\documentclass[11pt,english]{article}
\usepackage[T1]{fontenc}
\usepackage[utf8]{inputenc}
\usepackage{float}
\usepackage{amsmath}
\usepackage{amsthm}
\usepackage{amssymb}
\usepackage[bookmarks=false,
 breaklinks=false,pdfborder={0 0 1},backref=false,colorlinks=false]
 {hyperref}

\makeatletter

\providecommand{\tabularnewline}{\\}
\floatstyle{ruled}
\newfloat{algorithm}{tbp}{loa}
\providecommand{\algorithmname}{Algorithm}
\floatname{algorithm}{\protect\algorithmname}

\usepackage{fullpage}
\usepackage{algpseudocodex}
\algrenewcommand\algorithmicrequire{\textbf{Input:}}
\algrenewcommand\algorithmicensure{\textbf{Output:}}

\makeatother

\theoremstyle{plain}
\newtheorem{thm}{\protect\theoremname}
\newtheorem{lem}[thm]{\protect\lemmaname}
\newtheorem{prop}[thm]{\protect\propositionname}
\theoremstyle{definition}
\newtheorem{defn}[thm]{\protect\definitionname}
\theoremstyle{remark}
\newtheorem{claim}[thm]{\protect\claimname}
\usepackage{babel}
\providecommand{\claimname}{Claim}
\providecommand{\definitionname}{Definition}
\providecommand{\lemmaname}{Lemma}
\providecommand{\propositionname}{Proposition}
\providecommand{\theoremname}{Theorem}

\begin{document}
\global\long\def\rk{\operatorname{rk}}%

\global\long\def\cl{\operatorname{cl}}%

\global\long\def\dist{\operatorname{dist}}%

\global\long\def\OPT{\operatorname{OPT}}%

\global\long\def\ind{\mathbf{1}}%

\global\long\def\cI{\mathcal{I}}%

\global\long\def\cB{\mathcal{B}}%

\global\long\def\src{\mathsf{s}}%

\global\long\def\snk{\mathsf{t}}%
\allowdisplaybreaks
\title{A Faster Scaling Algorithm for Maximum Weight Matroid Intersection}
\author{
  Alina Ene\thanks{Department of Computer Science, Boston University, \texttt{aene@bu.edu}. This work was supported in part by a grant of access to OpenAI models through the ChatGPT for Academic Researchers program.} \and
  Huy L. Nguyen\thanks{Khoury College of Computer Sciences, Northeastern University, \texttt{hu.nguyen@northeastern.edu}. This work was supported in part by NSF CCF 2311649 and a gift from Apple Inc.}
}
\date{}
\maketitle
\begin{abstract}
We consider the maximum weight matroid intersection problem. In this
problem, we are given two matroids $M_{1}=(E,\mathcal{I}_{1})$ and
$M_{2}=(E,\mathcal{I}_{2})$ and integer weights $w(e)$ on the elements
$e\in E$, and the goal is to find a maximum-weight common independent
set. We give a randomized algorithm that constructs an optimal solution
using $\widetilde{O}(nr\log W)$ queries to the independence oracles
of the matroids, where $n$ is the size of the ground set $E$, $r$
is the minimum rank of the two matroids, and $W$ is the maximum weight.
\end{abstract}

\section{Introduction}

In the maximum weight matroid intersection problem, we are given two
matroids $M_{1}=(E,\cI_{1})$ and $M_{2}=(E,\cI_{2})$ and integer
weights $w(e)$ on the elements $e\in E$, and the goal is to find
a maximum-weight common independent set. We assume that the matroids
are given via independence oracles that take as input a set $S\subseteq E$
and return whether the set is independent in the respective matroid,
and the goal is to obtain an exact optimal solution using as few independence
queries as possible. The problem is a generalization of the classical
maximum weight bipartite matching problem and it is a cornerstone
problem in combinatorial optimization. The problem has received significant
attention, leading to the development of many algorithms for both
unweighted instances \cite{ad1971,lawler1975,edmonds1979,cunningham1986,lsw2015,nguyen19,clssw,bbmn,blikstad2021}
and weighted instances \cite{lawler1975,it1976,fujishige1977,frank1981,bcg1986,fz1995,si1995,lsw2015,hkk2019}.
Let $n$ be the size of the ground set $E$, $r$ be the minimum rank
of the two matroids, and $W$ be the maximum weight. For a long time,
the best known bounds in the weighted case for weakly polynomial algorithms
came from scaling algorithms \cite{fz1995,si1994,si1995,gx1996} until
the seminal work by \cite{lsw2015} using the entirely different approach
of a cutting plane method. The work \cite{lsw2015} uses $\widetilde{O}(n^{2}\log W)$
queries. However, this seems to be the limit of the current cutting
plane approach: for problems in $n$ dimensions, cutting plane methods
generally require $\Omega(n)$ cutting planes to make any progress.
Current implementation of the separation oracle call requires $\Omega(n)$
independence queries, leading to $\Omega(n^{2})$ queries overall.
However, for the combinatorial approach, there is no such barrier
and it is entirely possible to improve further. Indeed, for the special
case of unweighted instances, subsequent works by \cite{clssw,nguyen19}
showed that using binary search, one can obtain algorithms using only
$\widetilde{O}(nr)$ queries. The bound for the unweighted case was
then improved further to the state of the art result by \cite{blikstad2021}
using $\widetilde{O}(nr^{3/4})$ queries. Despite these improvements
in the unweighted case, their techniques unfortunately do not generalize
to the weighted case and the state of the art result for weighted,
weakly polynomial algorithms remains that of \cite{lsw2015}. In this
work, we give the first improvement over \cite{lsw2015} for weighted,
weakly polynomial algorithms using $\widetilde{O}(nr\log W)$ queries.
The prior state of the art results for various settings and our result
are summarized in Table~\ref{tab:bounds}. 

\begin{table}[t]
\centering
\begin{tabular}{|c|c|c|c|}
\hline 
Regime & Independence queries & Guarantee & Reference\tabularnewline
\hline 
\hline 
Weighted, strongly polynomial & $O(nr^{2})$ & Deterministic & \cite{bcg1986}\tabularnewline
\hline 
Weighted, weakly polynomial & $\widetilde{O}(n^{2}\log W)$ & Randomized & \cite{lsw2015}\tabularnewline
\hline 
Weighted, pseudopolynomial & $\widetilde{O}(nr^{5/6}W)$ & Deterministic & \cite{hkk2019}+\cite{blikstad2021}\tabularnewline
\hline 
Weighted, pseudopolynomial & $\widetilde{O}(nr^{3/4}W)$ & Randomized & \cite{hkk2019}+\cite{blikstad2021}\tabularnewline
\hline 
Unweighted & $\widetilde{O}(nr^{3/4})$ & Randomized &  \cite{blikstad2021}\tabularnewline
\hline 
Weighted, weakly polynomial & $\widetilde{O}(nr\log W)$ & Randomized & \textbf{This work}\tabularnewline
\hline 
\end{tabular}

\caption{Exact algorithms for matroid intersection.}\label{tab:bounds}
\end{table}

\begin{thm}
\label{thm:main} There is a randomized algorithm for weighted matroid
intersection that constructs an exact solution using $O\bigl(nr\log n\sqrt{\log r}\log W\bigr)$
independence queries in expectation.
\end{thm}

Our algorithm builds on the classical scaling approach for maximum
bipartite matching and more generally for matroid intersection \cite{fz1995,si1994,si1995},
the classical primal-dual approach based on weight splitting \cite{frank1981,frank2008},
and several recent works for unweighted matroid intersection \cite{clssw,nguyen19,bbmn}.
We give an overview of our algorithm and techniques in the following
section.

\subsection{Algorithm overview}

\label{subsec:overview}

The overall algorithm is a scaling algorithm (see Section \ref{sec:main-algorithm}),
which is a generalization of the scaling algorithm for maximum weight
matching (the special case where both matroids are partition matroids).
Initially the weights are all zero and we initialize with an arbitrary
common basis, which trivially maximizes the weight. In scaling phases,
the algorithm reveals one bit of the weight at a time, from the most
significant to the least significant. Thus, for each item $e$ with
the current weight $c$, the new weight is $c'=2c+\beta(e)$ where
$\beta(e)\in\{0,1\}$ is the next bit of $w(e)$. At the beginning
of the phase, we assume that we have a weight splitting $p+q=c$ and
a basis $B_{0}$ that maximizes $p$ in $M_{1}$ and maximizes $q$
in $M_{2}$, and hence, $B_{0}$ is the $c$-maximum common basis.
The goal of the phase is to obtain a new weight splitting $p'+q'=2c+\beta$
and a new common basis $B$ that maximizes $p'$ in $M_{1}$ and $q'$
in $M_{2}$. Initially, $p'=2p$ and $q'=2q+\beta$. We have $B_{1}=B_{0}$
and $B_{2}$ is the $q'$-maximizing basis in $M_{2}$ with the maximum
overlap with $B_{0}$. Let $X=B_{1}\cap B_{2}$ be our initial common
independent set. The primal progress will be measured by $|X|$, reaching
optimum when $|X|=r$. Let $F_{1}(p)=\max_{B\in\cB_{1}}p(B),F_{2}(q)=\max_{B\in\cB_{2}}q(B),U(p,q)=F_{1}(p)+F_{2}(q)$,
where $\cB_{i}$ is the set of all bases of $M_{i}$. The dual progress
will be measured by $U(p',q')-c'(B_{0})$. It can be shown that this
gap is always nonnegative as we adjust $p'$, $q'$ while maintaining
$p'+q'=c'$. The gap starts at $k=r-|X|$ and only decreases over
the execution of the algorithm.

The algorithm tries to increase $|X|$ by finding an augmenting path
in the exchange graph for the current prices $p,q$ (see Section \ref{sec:exchange-graph}).
Let $p_{1}=p$ and $p_{2}=q$. Let $N_{i}(p_{i})$ be the matroid
whose bases are the $p_{i}$-maximum bases of $M_{i}$ and $\cI(N_{i})$
the collection of independent sets of $N_{i}=N_{i}(p_{i})$. The exchange
graph has a source $\src$, a sink $\snk$, and vertices corresponding
to items in $E$. There is an edge from $\src$ to $e\in E\setminus X$
if $X+e\in\cI(N_{1})$. There is an edge from $e\in E\setminus X$
to $\snk$ if $X+e\in\cI(N_{2})$. For $e\in E\setminus X$ and $x\in X$,
there is an edge $e\to x$ if $X-x+e\in\cI(N_{2})$ and there is an
edge $x\to e$ if $X-x+e\in\cI(N_{1})$. If there is a chordless $\src-\snk$
path then we can use the corresponding exchanges to increase $|X|$
while remaining independent in $N_{1},N_{2}$. The key observation
from prior works \cite{clssw,nguyen19} is that the cost of finding
paths using BFS is proportional to the depth of the BFS, not the number
of vertices/edges explored. Building on this observation, we divide
the execution of the algorithm into two stages, when $k>\sqrt{r}$
and when $k\leq\sqrt{r}$. Consider the first stage, $k>\sqrt{r}$.
In the unweighted case, there is always a short augmenting path of
length $O(r/k)$, leading to low cost augmentation. The reader might
recall a similar argument for matching algorithms. However, this is
no longer true in the weighted case.

The issue in the weighted case is that when $|X|$ is already maximum
in $\cI(N_{1})\cap\cI(N_{2})$ but $|X|<r$, we need to adjust the
price splitting. This is typically done by finding a cut $\{\src\}\cup A$
in the exchange graph with no arcs from $\{\src\}\cup A$ to $(E\setminus A)\cup\{\snk\}$.
We then adjust the prices $p\gets p-\ind_{A},q\gets q+\ind_{A}$ and
maintain $p+q=c'$. It can be shown that the dual gap $U(p,q)-c'(B_{0})$
decreases by $k=r-|X|$. If we use BFS to find $A$, it might need
to search very deep, costing many queries. The novel idea is to trade
the primal progress for the dual progress. We search until depth $4\sqrt{r}$.
Suppose no augmenting path is found but we have not found all vertices
reachable from $\src$ either. The odd layers of BFS contain vertices
in $E\setminus X$ and the even layers contain vertices in $X$. Let
$\ell$ be the even layer of smallest size $b\le|X|/(2\sqrt{r})<\sqrt{r}/2$.
We can remove from $X$ all vertices in layer $\ell$ and use the
set of vertices reachable from $\src$ through layer $\ell$ as $A$
for price update. $A\cup\{\src\}$ is not a cut (there can be new
augmenting paths via vertices that were in layer $\ell$) but the
number of disjoint paths is bounded by $b$. It can be shown that
price update improves the dual gap by at least $k-b$ at the cost
of lowering $|X|$ by $b<\sqrt{r}/2$. Since $k>\sqrt{r}$, the dual
gain $k-b$ is much larger than the primal loss at most $b$. See
Section \ref{sec:thin-layer-repair} for details.

When augmentation is possible with short paths, a variant of the BFS
algorithm of \cite{clssw} that truncates the search at depth $O(\sqrt{r})$
can be used to augment with low amortized cost. See Section \ref{sec:truncated-bfs}.

In the second stage $k\leq\sqrt{r}$, we no longer need to trade off
between the primal and the dual and just need to find up to $\sqrt{r}$
augmenting paths. The execution is divided into epochs: each epoch
consists of a sequence of price updates followed by a single augmentation
(see Section \ref{sec:multicut-search}). If augmentation is possible,
the previous work \cite{bbmn} for the unweighted problem gives an
algorithm for finding one augmenting path using $\widetilde{O}\left(n\sqrt{r}\right)$
queries. If augmentation is not possible, we need to perform a series
of price updates leading to the next augmentation. The previous work
\cite{bbmn} introduced a method to find a subset $F\setminus\{\src\}\subseteq A$
with paths from $\src$ to all of $F$ as well as a categorization
of $E\setminus X$ into heavy and light depending on the number of
neighbors in $X\setminus F$ (neighbors of $\snk$ are also considered
heavy). We adapt this framework to maintain $F$ through a sequence
of price updates with no augmentation in between. An important observation
is that the price updates do not affect the paths from $\src$ to
$F$ so $F$ can be maintained and expanded through a sequence of
price updates. We also use the method of \cite{bbmn} to extend from
$F$ to a valid cut, enabling a price update. To optimize the cost,
the threshold for heavy/light needs to be set based on the number
of price updates, which we do not know a priori. We resolve this issue
using the doubling trick. The epoch is divided into blocks. Block
$i$ allows $2^{i}$ price updates and the heavy/light threshold is
set accordingly. We can index the epoch using the value $k=r-|X|$.
In epoch $k$, if there are $t_{k}$ price updates before the augmentation,
the expected cost of the algorithm is $\widetilde{O}(n\sqrt{r(t_{k}+1)})$.
Since the initial dual gap is at most $r$ and it decreases by $k$
after every price update, we have $\sum_{k}kt_{k}\le r$, leading
to the following bound using Cauchy-Schwarz: $\sum_{k}\sqrt{t_{k}}\le\sqrt{(\sum_{k}kt_{k})(\sum_{k}1/k)}=O(\sqrt{r\ln r})$.
Thus the total cost of the second stage is $\widetilde{O}(nr)$ queries.

\textbf{Acknowledgments:} We used GPT 6 Astra to develop the ideas,
write the initial proofs, and check correctness. The authors rewrote
and verified all of the proofs.

\section{Preliminaries}

\subsection{Matroid definitions and facts}

\label{sec:notation} A \emph{matroid} $M=(E,\cI)$ consists of a
finite ground set $E$ and a nonempty family $\cI\subseteq2^{E}$
of independent sets with the following properties:
\begin{itemize}
\item The empty set is independent, i.e. $\varnothing\in\mathcal{I}$.
\item Hereditary property: if $J\in\mathcal{I}$ and $I\subseteq J$ then
$I\in\mathcal{I}$.
\item Augmentation property: if $I,J\in\mathcal{I}$ and $|I|<|J|$, there
is an element $e\in J\setminus I$ such that $I\cup\{e\}\in\mathcal{I}$.
\end{itemize}
We assume that we are given access to the matroid via an independence
oracle that takes as input a set $S$ and it returns whether $S\in\cI$.

The \emph{bases} $\cB$ of $M$ are the inclusion-maximal independent
sets. All bases have the same cardinality.

For a set $S\subseteq E$, its \emph{rank} $\rk_{M}(S)$ and \emph{closure}
(or \emph{span}) $\cl_{M}(S)$ are defined as follows:
\begin{align*}
\rk_{M}(S) & =\max\left\{ \left|I\right|\colon I\subseteq S,\ I\in\cI\right\} ,\\
\cl_{M}(S) & =\{e\in E:\rk_{M}(S\cup\{e\})=\rk_{M}(S)\}.
\end{align*}
We also let $\rk(M):=\rk_{M}(E)$. We say that $S$ spans $T$ if
$T\subseteq\cl_{M}(S)$.

A \emph{circuit} is an inclusion-minimal dependent set. If $I$ is
independent and $I\cup\{e\}$ is dependent, its unique circuit is
denoted $C_{M}(e,I)$ and called the fundamental circuit of $e$ relative
to $I$.

The \emph{restriction} $M|H$ of $M$ to $H\subseteq E$ is the matroid
with ground set $H$ and independent sets $\{I\in\mathcal{I}:I\subseteq H$\}.
The \emph{contraction} $M/H$ has ground set $E\setminus H$ and rank
function 
\[
\rk_{M/H}(S)=\rk_{M}(S\cup H)-\rk_{M}(H).
\]
If $H$ is independent, $S$ is independent in $M/H$ exactly when
$S\cup H$ is independent in $M$.

The rank-$r$ truncation of a matroid $M$ is the matroid whose independent
sets are the independent sets of $M$ of size at most $r$.

\textbf{Greedy algorithm.} Given a weight function $f$, we can find
an $f$-maximum basis in a matroid $M$ using the Greedy algorithm:
we consider the elements in non-increasing order according to the
weight $f$ and we add each element if doing so preserves independence.
The optimality of the Greedy algorithm implies that a basis $B$ is
a maximum $f$-weight basis if and only if $f(x)-f(e)\geq0$ for every
$x\in B$ and $e\notin B$ such that $(B\setminus\{x\})\cup\{e\}\in\cI$.
We refer to $f(x)-f(e)$ as the slack of the exchange.

We can also use the Greedy algorithm to extend a given independent
set $I$ to a basis containing $I$ with the largest $f$-weight among
such bases: we start with $I$ and run the Greedy algorithm on the
remaining elements sorted in non-increasing $f$-weight order. Given
two weight functions $f$ and $g$, we can modify the Greedy algorithm
so that it breaks ties according to $g$. More precisely, we order
the elements in non-increasing $f$-weight, breaking ties in favor
of larger $g$-weight.

\textbf{Notation.} Throughout, we let $S+e=S\cup\{e\}$, $S-e=S\setminus\{e\}$,
and $S-T=S\setminus T$. The \emph{symmetric difference }of $S$ and
$T$ is $S\triangle T=(S\setminus T)\cup(T\setminus S)$. For any
function $f:E\to\mathbb{R}$ let $f(S)=\sum_{e\in S}f(e)$, including
$f(\varnothing)=0$. Function identities and inequalities are pointwise:
$p+q=c$ means $p(e)+q(e)=c(e)$ for every $e$. 

\subsection{Reduction to finding a common basis in two matroids of the same rank}

\label{sec:padding}

We consider the maximum-weight matroid intersection problem. We are
given two matroids $M_{1}=(E,\cI_{1})$ and $M_{2}=(E,\cI_{2})$ on
the same ground set $E$ and an integer weight function $w\colon E\to\mathbb{Z}$,
and the goal is to find a common independent set $I\in\cI_{1}\cap\cI_{2}$
of maximum weight $w(I):=\sum_{e\in I}w(e)$. We let $r:=\min\left\{ \rk(M_{1}),\rk(M_{2})\right\} $
be the minimum rank of the two matroids, $n:=\left|E\right|$ be the
size of the ground set, and $W=\max_{e\in E}w(e)$ be the maximum
weight. We may assume that the weights are non-negative, since we
can remove any element with negative weight.

In this section, we show that the maximum-weight matroid intersection
problem can be reduced to the problem of finding a maximum-weight
common base in two matroids that have a common base of size $r$.
Let $Z_{{\rm dum}}$ be a set of $r$ new elements, each of which
has zero weight; we refer to these elements as dummy elements. For
$i\in\{1,2\}$, we replace the matroid $M_{i}=(E,\cI_{i})$ with the
matroid $\widetilde{M}_{i}=(E\cup Z_{{\rm dum}},\widetilde{\cI}_{i})$
where
\[
\widetilde{\cI}_{i}=\{S\subseteq E\cup Z_{{\rm dum}}:|S|\le r,\ S\setminus Z_{{\rm dum}}\in\cI_{i}\}.
\]

We can verify that $\widetilde{M}_{i}$ is a matroid and it has rank
$r$. We can readily implement the independence oracle for $\widetilde{M}_{i}$
using the oracle for $M_{i}$: given a set $S$, we call the latter
on $S\setminus Z_{{\rm dum}}$ if $\left|S\right|\leq r$, and we
return not independent if $\left|S\right|>r$. The set $Z_{{\rm dum}}$
is a common basis for $\widetilde{M}_{1}$ and $\widetilde{M}_{2}$.
Additionally, every original common independent set $S\in\cI_{1}\cap\cI_{2}$
extends to a common basis in $\widetilde{\cI}_{1}\cap\widetilde{\cI}_{2}$
of the same weight by adding $r-\left|S\right|$ elements of $Z_{{\rm dum}}$
to $S$. Conversely, every common basis in $\widetilde{\cI}_{1}\cap\widetilde{\cI}_{2}$
can be mapped to a common independent set in $\cI_{1}\cap\cI_{2}$
of the same weight by removing the dummy elements. The new ground
set $E\cup Z_{{\rm dum}}$ has size $n+r\leq2n$.

Our algorithm finds a maximum-weight common basis in $\widetilde{\cI}_{1}\cap\widetilde{\cI}_{2}$,
which we then map to a common independent set in the original matroids
by removing the dummy elements $Z_{\mathrm{dum}}$. To simplify notation,
in the following, we relabel the new ground set and matroids as $E$
and $M_{i}=(E,\cI_{i})$, and we relabel the size of the new ground
set as $n$.

\subsection{Weight-splitting formulation}

\label{sec:query-optimal-face}

As discussed above, we consider the problem of finding a maximum-weight
common basis in two matroids $M_{1}=(E,\cI_{1})$ and $M_{2}=(E,\cI_{2})$.
For a weight function $f$ and a matroid $M$, we refer to a basis
$B\in\arg\max_{S\in\cB(M)}f(S)$ as an $f$-maximum basis of $M$.
Following previous work \cite{frank1981,frank2008}, we use the following
primal-dual weight-splitting formulation of the problem. Let $\cB_{i}$
be the family of bases of $M_{i}$. Let $c:E\to\mathbb{Z}$ be an
integer weight function. For integer-valued functions $p$ and $q$
on the ground set $E$, we say that $p,q$ is a weight-splitting of
$c$ if $c=p+q$. We refer to $p$ and $q$ as prices. Let
\[
F_{1}(p)=\max_{B\in\cB_{1}}p(B),\qquad F_{2}(q)=\max_{B\in\cB_{2}}q(B),\qquad U(p,q)=F_{1}(p)+F_{2}(q).
\]
As shown in \cite{frank1981,frank2008}, we have
\[
\max_{B\in\cB_{1}\cap\cB_{2}}c(B)=\min_{p,q\colon E\to\mathbb{Z},p+q=c}U(p,q)
\]
Note that, for any weight-splitting $c=p+q$, $U(p,q)$ is an upper
bound on the weight of a $c$-maximum common basis, and we refer to
$U(p,q)$ as a dual upper bound on the common-basis optimum. 

Consider any weight-splitting $c=p+q$, and let $p_{1}=p$ and $p_{2}=q$.
For $i\in\{1,2\}$, let $N_{i}(p_{i})$ denote the matroid whose bases
are the $p_{i}$-maximum bases of $M_{i}$; this is a matroid \cite[Proposition 2]{frank2008}.
Note that, if $B$ is a common basis of $N_{1}(p_{1})$ and $N_{2}(p_{2})$,
we have $p(B)=F_{1}(p_{1})$ and $q(B)=F_{2}(p_{2})$ and thus $B$
is a $c$-maximum common basis. This observation forms the basis of
our algorithm. More precisely, the algorithm maintains a weight-splitting
$c=p_{1}+p_{2}$ and a common independent set $X\in\cI(N_{1}(p_{1}))\cap\cI(N_{2}(p_{2}))$.
If $\left|X\right|=r$, $X$ is an optimal solution and the algorithm
terminates. Otherwise, the algorithm tries to increase the size of
$X$ and it updates the weight-splitting if it fails to do so.

In the following, for brevity, we refer to the matroids $N_{1}(p_{1})$
and $N_{2}(p_{2})$ as the \emph{price-optimal matroids} (for the
current weight-splitting $c=p_{1}+p_{2}$). Since the algorithm will
work with these matroids, we need to test independence in these matroids.
We now describe how to do so. In addition to the common independent
set $X\in\cI(N_{1}(p_{1}))\cap\cI(N_{2}(p_{2}))$, the algorithm also
maintains a basis $B_{i}\in\arg\max_{B\in\cB_{i}}p_{i}(B)$ such that
$X\subseteq B_{i}$ for each $i\in\{1,2\}$. For a given set $T$
and element $e$, the algorithm needs to check whether $(X\setminus T)\cup\{e\}$
is independent in $N_{i}(p_{i})$. Algorithm \ref{alg:opt-query}
performs this check using a single query to the independence oracle
$\operatorname{InputInd}_{i}$ of the original matroid $M_{i}$ (before
adding the dummy elements). For a value $\lambda$ and set $S$, we
let $S_{>\lambda}=\{e\in S:p_{i}(e)>\lambda\}$ and $S_{=\lambda}=\{e\in S:p_{i}(e)=\lambda\}$.

\begin{algorithm}[tbh]
\begin{algorithmic}[1]
\Function{Probe}{$i,T,e$}
  \State $\lambda\gets p_i(e)$;
    $H\gets(B_i)_{>\lambda}\cup(X\setminus T)_{=\lambda}\cup\{e\}$.
  \If{$|H|>r$}
    \State \Return false
  \EndIf
  \State \Return $\operatorname{InputInd}_i(H\setminus Z_{\rm dum})$
\EndFunction
\end{algorithmic}

\caption{Algorithm for checking whether $(X\setminus T)\cup\{e\}$ is independent
in $N_{i}(p_{i})$. The algorithm has access to $X$, $p_{i}$, and
a base $B_{i}$ of $N_{i}(p_{i})$ that contains $X$. }\label{alg:opt-query}
\end{algorithm}

The following lemma shows that the algorithm is correct. 
\begin{lem}
[One original query]\label{lem:one-query} Let $B$ be a $p_{i}$-maximum
basis of $M_{i}$, let $I$ be independent in $N_{i}(p_{i})$, and
let $e\notin I$. For $\lambda=p_{i}(e)$, 
\[
I+e\in\cI(N_{i}(p_{i}))\quad\Longleftrightarrow\quad B_{>\lambda}\cup I_{=\lambda}\cup\{e\}\in\cI_{i}.
\]
\end{lem}

\begin{proof}
To simplify notation, we drop the subscript $i$ and let $M=M_{i}$,
$p=p_{i}$, and $N=N_{i}(p_{i})$. Let $H=\{e'\in E:p(e')>\lambda\}$
and $J=I_{=\lambda}\cup\{e\}$. Let $C$ be any $p$-maximum basis
of $M$. The optimality of $C$ implies that $C_{>\lambda}$ is a
basis of $M|H$. Thus $C_{>\lambda}$ and $B_{>\lambda}$ have the
same cardinality and span. Since $J\cap H=\varnothing$, it follows
that
\[
B_{>\lambda}\cup J\in\cI(M)\quad\Longleftrightarrow\quad C_{>\lambda}\cup J\in\cI(M).
\]
Suppose first that $I+e$ is independent in $N$. Let $D$ be a $p$-maximum
basis containing $I+e$. Then $D_{>\lambda}\cup J\subseteq D$ is
independent, and the above equivalence implies that $B_{>\lambda}\cup J$
is independent.

Conversely, suppose that $B_{>\lambda}\cup J$ is independent. Let
$C$ be a $p$-maximum basis containing $I$. If $e\in C$, then $I+e$
is already contained in a $p$-maximum basis. Otherwise, let $F$
be the fundamental circuit of $e$ with respect to $C$. For every
$x\in F\setminus\{e\}$, the set $C-x+e$ is a basis, so optimality
of $C$ implies $p(x)\ge p(e)=\lambda$. The set $C_{>\lambda}\cup J$
is independent by the above equivalence and therefore cannot contain
$F$. Hence there is an $x\in F\setminus\{e\}$ outside $C_{>\lambda}\cup I_{=\lambda}$.
It follows that $p(x)=\lambda$ and $x\notin I$. Thus $C-x+e$ is
a $p$-maximum basis containing $I+e$, and thus $I+e$ is independent
in $N$.
\end{proof}

As noted above, the algorithm maintains a common independent set $X$
of both $M_{1}$ and $M_{2}$, and a maximum-price basis $B_{i}$
of $M_{i}$ containing $X$ for $i=1,2$. Whenever $X$ is updated,
we can construct $B_{i}$ by extending $X$ using the Greedy algorithm
as discussed in Section \ref{sec:notation}. This takes $O(n)$ original
queries.

\section{The main algorithm}

\label{sec:main-algorithm}

Our main algorithm is a bit scaling algorithm described in Algorithm
\ref{alg:weighted-scaling}. 

\begin{algorithm}[tbh]
\begin{algorithmic}[1]
\Require Two matroid independence oracles on $E$ and integer weights $w$.
\Ensure A maximum-weight common independent set.
\State Discard all negative-weight elements from both matroids.
\State $W\gets\max(\{0\}\cup\{w(e):e\in E\})$;
       $r\gets\min\{\rk(M_1),\rk(M_2)\}$, found greedily.
\If{$r=0$ or $W=0$}
  \State \Return $\varnothing$
\EndIf
\State Pad by $Z_{\rm dum}$ and truncate both matroids to rank $r$
       as in Section~\ref{sec:padding}.
\State $w(e)\gets0$ for $e\in Z_{\rm dum}$;
       $B\gets Z_{\rm dum}$; $c,p,q\gets0$.
\State $\ell\gets\lceil\log_2(W+1)\rceil$.
\For{$b=\ell-1,\ell-2,\ldots,0$}
  \State $\beta(e)\gets\lfloor w(e)/2^b\rfloor\bmod2$ for every $e\in E$.
  \State $(B,p,q)\gets\Call{RefineBit}{B,p,q,\beta}$.
  \State $c\gets 2c+\beta$.
  \Comment{$p+q=c$ and $B$ is optimal for both prices}
\EndFor
\State \Return $B\setminus Z_{\rm dum}$
\end{algorithmic}

\caption{Main algorithm for maximum weight matroid intersection using bit
scaling}\label{alg:weighted-scaling}
\end{algorithm}

We review the properties of weight adjustment in the standard bit
scaling algorithm, which are consequences of Frank's weight adjustment
formulas. For completeness, we also include their direct proofs.
\begin{prop}
[Frank's weight adjustment formulas]\cite[Proposition 3]{frank2008}
Let $M$ be a matroid on $E$, let $f:E\to\mathbb{Z}$ be integral,
and let $N$ be the matroid whose bases are the $f$-maximum bases
of $M$. For any weight function $a$, write $F_{M}(a)=\max_{B\in\mathcal{B}(M)}a(B)$.
Then every $H\subseteq E$ satisfies 
\begin{align}
F_{M}(f+\ind_{H}) & =F_{M}(f)+\rk_{N}(H),\label{eq:frank-plus}\\
F_{M}(f-\ind_{H}) & =F_{M}(f)-\rk(M)+\rk_{N}(E\setminus H).\label{eq:frank-minus}
\end{align}
\end{prop}

The bit scaling algorithm works in phases. In each phase, the algorithm
exposes the next binary digit of every input weight, from the most
significant to the least significant, for a total of $\lceil\log_{2}(W+1)\rceil$
phases. At the beginning of each scaling phase, we start with the
weight vector $c$, which is the integer prefix of bits already exposed
in previous phases. We also have a weight splitting $p+q=c$ and a
common basis $B_{0}$ that is both a $p$-maximum basis of $M_{1}$
and a $q$-maximum basis of $M_{2}$, and hence, a maximum common
independent set for weight $c$. For initialization, the algorithm
starts with $B_{0}$ being the dummy basis, weight vector $c=0$ and
the weight splitting $p=q=0$. In each phase, the algorithm exposes
one new bit $\beta\in\{0,1\}^{E}$ and lets 
\[
c'=2c+\beta,\qquad p_{0}=2p,\qquad q_{0}=2q+\beta.
\]

The bulk of the phase, implemented in the function \textsc{RefineBit}
in Section \ref{sec:near-nr-bound}, is to find a new weight splitting
$p'+q'=c'$ and a new basis $B$ that is $p'$-maximum in $M_{1}$,
$q'$-maximum in $M_{2}$ and therefore, a $c'$-maximum common basis.
The algorithm keeps $B_{1}=B_{0}$, and chooses a $q_{0}$-maximum
basis $B_{2}$ with maximum overlap with $B_{0}$. The greedy algorithm,
breaking equal-price ties in favor of $B_{0}$, makes this choice
in $O(n)$ queries. Initialize $X=B_{0}\cap B_{2}$ and define the
initial gap 
\[
\Gamma=U(p_{0},q_{0})-c'(B_{0}).
\]
Throughout the phase, $\Gamma$ denotes this fixed initial gap. The
changing \emph{deficit} is $k=r-|X|$; it counts how many further
elements are needed for a common basis.
\begin{lem}
[Gap equals initial deficit]\label{lem:warm} The initialization
satisfies 
\[
\Gamma=r-|X|=|B_{2}\setminus B_{0}|\le r.
\]
\end{lem}

\begin{proof}
Since $q$ is integral, sorting by $2q+\beta$ refines the sorting
by $q$. Thus $B_{2}$ is also a maximum-$q$ basis, and it maximizes
$\beta$ in the old price-optimal matroid $N_{2}(q)$. Both $B_{0}$
and $B_{2}$ are bases there.

No element $e\in B_{0}\setminus B_{2}$ has $\beta(e)=1$. Otherwise
basis exchange would allow insertion of $e$ into $B_{2}$ while removing
an element of $B_{2}\setminus B_{0}$: this either improves $\beta(B_{2})$
or preserves $\beta(B_{2})$ and improves the size of the overlap
$B_{2}\cap B_{0}$. Likewise, no element $e\in B_{2}\setminus B_{0}$
has $\beta(e)=0$, using the basis-exchange property in the opposite
direction. Hence every removed element has bit zero and every inserted
element has bit one. Since $q(B_{2})=q(B_{0})$ and $B_{0}$ is $p_{0}$-maximum,
\[
\Gamma=q_{0}(B_{2})-q_{0}(B_{0})=\beta(B_{2})-\beta(B_{0})=|B_{2}\setminus B_{0}|.
\]
\end{proof}

 The algorithm maintains $X$ independent in $\cI(N_{1}(p))\cap\cI(N_{2}(q))$
and two optimal bases $B_{i}\supseteq X$. A set $A\subseteq E$ is
called a \emph{min--max witness for $X$}, if it satisfies 
\begin{equation}
\rk_{N_{1}}(E\setminus A)+\rk_{N_{2}}(A)=|X|.\label{eq:minmax}
\end{equation}

This is a certificate that $X$ has the maximum cardinality in $\cI(N_{1}(p))\cap\cI(N_{2}(q))$
. If $|X|<r$, it shows that there is no common basis that simultaneously
maximizes $p$ in $M_{1}$ and $q$ in $M_{2}$. In this case, we
need to update the prices as follows. 
\begin{equation}
p'\gets p-\ind_{A},\qquad q'\gets q+\ind_{A}.\label{eq:update}
\end{equation}

\begin{lem}
[Preservation after price update]\label{lem:unit} Old $p$-maximum
basis $B_{1}$ containing $X$ is also a $p'$-maximum basis. Similarly,
old $q$-maximum basis $B_{2}$ is a $q'$-maximum basis. The set
$X$ remains independent in both new price-optimal matroids $N_{1}(p')$
and $N_{2}(q')$, and 
\[
U(p',q')=U(p,q)-(r-|X|).
\]
\end{lem}

\begin{proof}
First, $X\setminus A$ is independent in $N_{1}(p)$ and $X\cap A$
is independent in $N_{2}(q)$, so 
\[
\rk_{N_{1}(p)}(E\setminus A)\ge|X\setminus A|,\qquad\rk_{N_{2}(q)}(A)\ge|X\cap A|.
\]
The two lower bounds sum to $|X|$, and the two ranks also sum to
$|X|$ by~(\ref{eq:minmax}). Neither inequality can therefore be
strict. Thus 
\[
\rk_{N_{1}(p)}(E\setminus A)=|X\setminus A|,\qquad\rk_{N_{2}(q)}(A)=|X\cap A|.
\]
Equivalently, $X\setminus A$ is a basis of $N_{1}(p)|(E\setminus A)$
and $X\cap A$ is a basis of $N_{2}(q)|A$. Because $B_{1}$ is a
basis of $N_{1}(p)$ containing $X$, 
\[
|X\setminus A|\le|B_{1}\setminus A|\le\rk_{N_{1}(p)}(E\setminus A)=|X\setminus A|.
\]
Applying the analogous argument to $B_{2}\cap A$ gives $|B_{1}\setminus A|=|X\setminus A|$
and $|B_{2}\cap A|=|X\cap A|$.

A basis is $p$-maximum if and only if every feasible single exchange
removing $x$ and inserting $y$ has nonnegative slack $p(x)-p(y)$.
For $B_{1}$, the only slack decreased by the update (i.e. $p'(x)-p'(y)<p(x)-p(y)$)
removes $x\in B_{1}\cap A$ and inserts $y\notin B_{1}\cup A$. If
the old slack were zero, $B_{1}-x+y$ would be an old $p$-maximum
basis with more than $|B_{1}\setminus A|=\rk_{N_{1}(p)}(E\setminus A)$
elements outside $A$, which is impossible. The integral slack $p(x)-p(y)$
is therefore at least one and thus, $p'(x)-p'(y)\ge0$. The argument
for $B_{2}$ is symmetric, with $x\in B_{2}\setminus A$ and $y\in A\setminus B_{2}$.
Thus both bases stay optimal and still contain $X$.

Evaluating the new optima on these same bases gives 
\[
U(p',q')-U(p,q)=-|B_{1}\cap A|+|B_{2}\cap A|=-r+|X\setminus A|+|X\cap A|=-(r-|X|).
\]
\end{proof}

The above Lemma suggests the overall strategy. The algorithm tries
to increase the size of $X$ iteratively. When $X$ has the maximum
cardinality in $\cI(N_{1}(p))\cap\cI(N_{2}(q))$ but $\left|X\right|<r$,
we adjust the prices as above, which reduces the gap $U(p,q)-c'(B_{0})$.
Our measure of progress is in terms of $|X|$ and the gap $U(p,q)-c'(B_{0})$,
both of which can only be improved a finite number of times so the
algorithm eventually terminates. Section \ref{sec:exchange-graph}
will describe the exchange graph, which is the mechanism to increase
the size of $X$ if possible. Section \ref{sec:truncated-bfs} describes
a truncated variant of BFS on the exchange graph to find a sequence
of augmenting paths on the exchange graph with amortized cost. Section
\ref{sec:thin-layer-repair} describes how to trade off a small decrease
in $|X|$ for a larger decrease in $U(p,q)-c'(B_{0})$, allowing us
to perform a price update more cheaply. Section \ref{sec:near-nr-bound}
puts together these ingredients to implement \textsc{RefineBit}, the
procedure for one bit scaling phase.

\section{The exchange graph }

\label{sec:exchange-graph}

In this section, we define the exchange graph used by the algorithm
for increasing the size of the current common independent set $X$
by finding an augmenting path in the graph if one exists. We state
several results from prior work for traversing the graph in search
of an augmenting path.

\textbf{The exchange graph.} In this section, ``inside'' means
in $X$, and ``outside'' means in $E\setminus X$. The source $\src$
and sink $\snk$ are artificial vertices not in $E$ and are always
distinct. For fixed matroids $N_{1},N_{2}$ and a common independent
set $X$, the graph $G_{X}$ has vertices $E\cup\{\src,\snk\}$ and
the following arcs:
\begin{align*}
\src\to e & \quad\text{if }X+e\in\cI(N_{1}),\quad e\notin X,\\
e\to\snk & \quad\text{if }X+e\in\cI(N_{2}),\quad e\notin X,\\
x\to e & \quad\text{if }X-x+e\in\cI(N_{1}),\quad x\in X,\ e\notin X,\\
e\to x & \quad\text{if }X-x+e\in\cI(N_{2}),\quad x\in X,\ e\notin X.
\end{align*}
We define the length of a directed path to be the number of arcs on
the path. Let $\dist_{G_{X}}(u,v)$ be the shortest directed-path
length in $G_{X}$, with value $+\infty$ if no such path exists;
we drop the subscript $X$ when the graph is clear. We refer to the
set $\{v:\dist_{G_{X}}(\src,v)=j\}$ of vertices at distance $j$
from the source $\src$ as the source layer of index $j$. Breadth-first
search (BFS) explores these layers in increasing order. Reachable
vertices outside $X$ have odd distance from $\src$ and reachable
vertices inside $X$ have even distance from $\src$; a finite source--sink
distance is even. 

\textbf{Augmenting paths.} We say that a directed path is \emph{chordless}
if it has no forward arc skipping at least one intermediate path vertex.
In particular, a shortest path is chordless. Given a chordless $\src-\snk$
path $P$, we can augment $X$ as follows. Let $V_{E}(P)$ denote
the ground-set vertices on $P$, excluding the terminals $\src,\snk$.
We can obtain a larger common independent set by considering the set
$X\triangle V_{E}(P)$, where $\triangle$ denotes the symmetric difference.
The following proposition, which follows from previous work \cite{clssw,bbmn},
shows that we can find a larger common independent set in $\cI(N_{1})$
and $\cI(N_{2})$ than $X$ if and only if there is a chordless augmenting
path from $\src$ to $\snk$, which is a generalization of augmenting
paths for matching.
\begin{prop}
[augmentation and circuit facts] \label{prop:import-structural}
For any fixed pair of matroids and common independent set $X$: 
\begin{enumerate}
\item \cite[Lemma 5]{clssw} If $X+e$ is dependent in one matroid $N$,
then 
\[
C_{N}(e,X)=\{z\in X+e:(X+e)-z\in\mathcal{I}(N)\}.
\]
\cite[Claim 2.5]{bbmn} Thus, for nonempty $T\subseteq X$, $(X\setminus T)+e$
is independent exactly when there exists $x\in T$ with $X-x+e$ independent.
If $X+e$ is independent, it also holds that for nonempty $T\subseteq X$,
$(X\setminus T)+e$ is independent exactly when there exists $x\in T$
with $X-x+e$ independent. 
\item \cite[Claim 2.2]{bbmn} Every chordless $\src$--$\snk$ path produces
a common independent set $X\triangle V_{E}(P)$ of size $|X|+1$.
In particular, a shortest path suffices \cite[Lemma 6]{clssw}. Moreover,
$X$ is of maximum common cardinality if and only if $G_{X}$ contains
no $\src$--$\snk$ path \cite[Corollary 2.3]{bbmn}. 
\end{enumerate}
\end{prop}

\textbf{Closed separating sets.} A set $A\subseteq E$ is called a
\emph{closed separating set} if it contains every out-neighbor of
$\src$, contains no in-neighbor of $\snk$, and no exchange arc goes
from $A$ to $E\setminus A$.  If the graph does not contain an augmenting
path, one closed separating set can be found via a breadth-first-search
(BFS) from $\src$ in the exchange graph. Our algorithm uses a more
efficient implementation. The following lemma shows that any closed
separating set gives us a min-max witness for the dual update as in
equation (\ref{eq:minmax}). 
\begin{lem}
[A closed set is a min-max witness]\label{lem:closed} Every closed
separating set satisfies Equation (\ref{eq:minmax}). 
\end{lem}

\begin{proof}
For $e\in(E\setminus A)\setminus X$, $X+e$ is dependent in $N_{1}$,
since $e$ is not a source neighbor. Its fundamental circuit cannot
contain $x\in X\cap A$, since that would create an arc $x\to e$
leaving $A$. Hence $e$ is spanned by $X\setminus A$ in $N_{1}$.
Similarly every $e\in A\setminus X$ is spanned by $X\cap A$ in $N_{2}$.
It follows that the two ranks in~(\ref{eq:minmax}) are respectively
$\rk_{N_{1}}(E\setminus A)=|X\setminus A|$ and $\rk_{N_{2}}(A)=|X\cap A|$. 
\end{proof}

\textbf{Efficient access to the graph.} Among other routines, the
algorithm performs BFS to search the graph.\textbf{ } While the definition
allows an obvious way to test if an edge $x\to e$ or $e\to x$ exists
using one independence query, this is a query to the adjacency matrix
of the graph, which is very inefficient for implementing BFS. The
next proposition, which follows from previous work \cite{clssw,nguyen19,bbmn},
shows that one can efficiently find one incoming or outgoing neighbor
of $e\notin X$ using only a logarithmic number of queries. This is
similar to a query to the adjacency list, which allows for efficient
implementation of BFS.
\begin{prop}
[neighborhood-search interface]\label{prop:import-neighbor} Let
$I$ be independent in a matroid, $e\notin I$, and $\varnothing\ne T\subseteq I$.
The routine \textsc{FindExchange} (\cite[Algorithm 2]{clssw}, see
also \cite{nguyen19}) finds an $x\in T$ with $I-x+e$ independent,
or certifies there is none, using $O(\log(|T|+1))$ independence queries
\cite[Lemma 11]{clssw}. All its query sets have the form $(I\setminus T')+e$
with $T'\subseteq T$. In the exchange graph, \cite{bbmn}'s \textsc{InEdge}
and \textsc{OutEdge} apply this binary search to find an incoming
or outgoing neighbor of a vertex $e\notin X$ in a specified subset
of $X$ \cite[Claim 2.6]{bbmn}. Their cost is $O(\log(\sigma+1))$
queries, where $\sigma=|X|$. If \textsc{InEdge} needs to check if
there is an arc $\src\to e,$ it costs one additional query. If \textsc{OutEdge}
needs to check if there is an arc $e\to\snk,$ it costs one additional
query.
\end{prop}

\section{Truncated BFS algorithm}

\label{sec:truncated-bfs}

In this section, we describe a variant of the BFS algorithm in \cite[Algorithms 6-8]{clssw}
that only explores vertices within a given distance from the source
$\src$ in the exchange graph. \cite[Algorithms 6-8]{clssw} starts
from valid lower bounds for distances from $\src$ and repeatedly
augments along shortest paths. The adaptation in this section also
repeatedly augments along shortest paths but only until the shortest
$\src-\snk$ distance exceeds a prescribed value $d$. We first review
the monotonicity of the distances after augmentation.
\begin{prop}
[capped distance monotonicity]\cite[Lemma 9]{clssw} \label{prop:import-distance}
Let $P$ be a shortest $\src$--$\snk$ path in $G_{X}$, let $\ell=\dist_{G_{X}}(\src,\snk)$,
and augment to $X'=X\triangle V_{E}(P)$.  For every ground vertex
$v\in E$, 
\begin{align*}
\dist_{G_{X'}}(\src,v) & \ge\min\{\dist_{G_{X}}(\src,v),\ell\},\\
\dist_{G_{X'}}(v,\snk) & \ge\min\{\dist_{G_{X}}(v,\snk),\ell\}.
\end{align*}
Unreachable vertices have distance $+\infty$, with $\min\{+\infty,\ell\}=\ell$. 
\end{prop}

A consequence is that after the augmentation, the new $\src-\snk$
distance is at least the old value. Indeed, if a new $\src-\snk$
path had length $L<\ell$, choose any intermediate vertex $v$ on
it. We have $\dist_{G_{X'}}(\src,v)+\dist_{G_{X'}}(v,\snk)\le L$
and both $\dist_{G_{X'}}(\src,v),\dist_{G_{X'}}(v,\snk)\le\ell$.
The two inequalities in Proposition \ref{prop:import-distance} force
$\dist_{G_{X'}}(\src,v)\ge\dist_{G_{X}}(\src,v)$ and $\dist_{G_{X'}}(v,\snk)\ge\dist_{G_{X}}(v,\snk)$.
It follows that $\text{\ensuremath{\dist_{G_{X}}}(\ensuremath{\src},v)}+\dist_{G_{X}}(v,\snk)\le L<\ell$,
which contradicts the definition of $\ell=\dist_{G_{X}}(\src,\snk)$.

Next we describe an adaptation of the BFS in \cite[Algorithm 6]{clssw}
with two main changes: 1) we only explore until a given distance bound
$d$; and 2) we record a parent pointer whenever the algorithm discovers
a new vertex. These parent pointers can be used to trace back the
shortest path directly.

\textbf{Inputs, shared state, and outputs.} The main function \textsc{WarmSearch}
takes the current common independent set $X$, fixed split prices
$p,q$, optimal bases $B_{1},B_{2}$ containing $X$ in the two matroids,
an integer distance bound $d\ge2$, and a target cardinality $\tau$
with $|X|\le\tau\le r$. Throughout one call the prices are unchanged,
so $N_{1}(p),N_{2}(q)$ are fixed, although $X$ and $B_{1},B_{2}$
change after every augmentation. We assume $r\ge1$.

All helper procedures share the current $X,p,q,B_{1},B_{2}$ and a
vector $\mathbf{d}=(d_{v})_{v\in E}$ of finite lower bounds on the
distances from $\src$. During a layer search, $d_{v}$ becomes exact
when $v$ is discovered and remains a lower bound otherwise. The scalar
$d$ denotes the search horizon, while $\mathbf{d}$ denotes the label
vector. Let $p_{1}=p$ and $p_{2}=q$. A vertex $e\notin X$ has an
odd distance from $\src$ that is at least one; a vertex $x\in X$
has an even distance from $\src$ that is at least two. In a single
call to \textsc{LayerSearch}, $R\subseteq E$ is the set of discovered
vertices, $L_{j}$ is the completed distance-$j$ layer, and $\pi(v)$
is the recorded parent of discovered vertex $v$. These objects are
shared by the search helpers, and \textsc{LayerSearch} leaves its
final $R$ available to the main function \textsc{WarmSearch}. They
are rebuilt after every augmentation; only $\mathbf{d}$ is retained.
The set $\mathcal{T}$ contains the current neighbors of $\snk$.
The symbol $\bot$ means that a requested exchange does not exist,
and $\varnothing$ is also used for an unused output certificate.

\textsc{WarmSearch} returns the updated $X,B_{1},B_{2}$ and a status
with certificate $\mathcal{C}$: \texttt{target} if $|X|=\tau$; \texttt{cut}
with a closed separating set $A\subseteq E$; or \texttt{horizon}
with layers $L_{1},\ldots,L_{d}$ of all vertices at distance $1,\ldots,d$
and no source--sink path of length at most $d$. The internal status
\texttt{path} carries a shortest source--sink path. For a path $P$,
$V_{E}(P)$ excludes its two terminals and $|P|$ below denotes its
number of arcs.

\textbf{Every oracle call.} In Algorithm~\ref{alg:opt-query}, $\operatorname{InputInd}_{i}(H)$
means a call to the original input independence oracle on $H$, before
padding. Price slices in \textsc{Probe} use side $i$ and the current
prices. Thus $\textsc{Probe}(i,T,e)$ tests $(X\setminus T)+e$ in
$N_{i}$ using the explicit one-query construction of Lemma~\ref{lem:one-query}.
Its preconditions are $T\subseteq X$ and $e\notin X$. For $T=\varnothing$
it is exactly the source test ($i=1$) or sink test ($i=2$). The
size test implements the padded rank-$r$ truncation.

The layer routine calls $\textsc{FindExchange}(N_{i},X,e,T)$ with
$e\notin X$ and $T\subseteq X$, exactly as in Proposition~\ref{prop:import-neighbor}.
It returns an $x\in T$ with $X-x+e$ independent in $N_{i}$, or
$\bot$ if no such $x$ exists. For $i=1$, this finds an incoming
arc $x\to e$; for $i=2$, it finds an outgoing arc $e\to x$. Every
internal oracle query of \textsc{FindExchange} has the form $(X\setminus T')+e$
with $T'\subseteq T$ and is answered by $\textsc{Probe}(i,T',e)$.
Thus the call uses the stored optimal basis $B_{i}$ and original
oracle. When $T=\varnothing$, the call returns $\bot$ without any
original oracle query.

\textbf{Building the next layer.} Algorithm~\ref{alg:warm-layer}
is called to compute the layer $L_{j}$ of the BFS from $\src$ with
$j\ge2$ and the previous layer complete. For odd $j$, we need to
find all $e\in(E\setminus X)\setminus R$ that are undiscovered and
reachable from $L_{j-1}$. To do so, we test all candidate $e$ with
$d_{e}=j$ using \textsc{FindExchange}. If no arc is found, we can
increase the lower bound $d_{e}$ by two. For even $j$, we need to
find all $e\in X$ that are undiscovered and reachable from $L_{j-1}$.
We start with the candidate set $Q=X\setminus R$. Iterate over all
$e\in L_{j-1}$ and use \textsc{FindExchange} to find arcs to $Q$.
Every time we find an arc to a new vertex, we add it to $L_{j}$ and
remove it from $Q$.

\begin{algorithm}[tbh]
\begin{algorithmic}[1]
\Require $j\ge2$, completed $L_{j-1}$, and the shared search state $R,\mathbf d,\pi$.
\Ensure Exact layer $L_j$; $R,\mathbf d,\pi$ updated on discovery.
\Function{NextLayer}{$j,L_{j-1}$}
  \State $L_j\gets\varnothing$.
  \If{$j$ is odd}
    \State $Q\gets\{e\in E\setminus(X\cup R):d_e\le j\}$.
    \For{each $e\in Q$}
      \State $x\gets\Call{FindExchange}{N_1,X,e,L_{j-1}}$.
      \If{$x\ne\bot$}
        \State $L_j\gets L_j\cup\{e\}$; $R\gets R\cup\{e\}$;
          $d_e\gets j$; $\pi(e)\gets x$.
      \Else
        \State $d_e\gets j+2$.
      \EndIf
    \EndFor
  \Else
    \State $Q\gets X\setminus R$.
    \For{each $e\in L_{j-1}$}
      \Loop
        \State $x\gets\Call{FindExchange}{N_2,X,e,Q}$.
        \If{$x=\bot$}
          \State \textbf{break}
        \EndIf
        \State $Q\gets Q\setminus\{x\}$;
          $L_j\gets L_j\cup\{x\}$; $R\gets R\cup\{x\}$.
        \State $d_x\gets j$; $\pi(x)\gets e$.
      \EndLoop
    \EndFor
  \EndIf
  \State \Return $L_j$
\EndFunction
\end{algorithmic}\caption{Explore one layer of the exchange graph as part of a BFS}\label{alg:warm-layer}
\end{algorithm}

\textbf{BFS and augmentation up to a bounded distance.} Algorithm~\ref{alg:warm-bfs}
scans the arcs from $\src$ and arcs to $\snk$ once for the current
$X$ and then builds layers in order. When an augmenting path is detected,
it completes the last odd layer first before returning an augmenting
path through one of the neighbors of $\snk$.

\begin{algorithm}[tbh]
\begin{algorithmic}[1]
\Function{LayerSearch}{$d$}
  \State $L_0\gets\{\src\}$; $L_1\gets\varnothing$; $\mathcal T\gets\varnothing$;
    clear all predecessor pointers.
  \For{each $e\in E\setminus X$}
    \If{$\Call{Probe}{1,\varnothing,e}$}
      \State $L_1\gets L_1\cup\{e\}$; $d_e\gets1$; $\pi(e)\gets\src$.
    \Else
      \State $d_e\gets\max\{d_e,3\}$.
    \EndIf
    \If{$\Call{Probe}{2,\varnothing,e}$}
      \State $\mathcal T\gets\mathcal T\cup\{e\}$.
    \EndIf
  \EndFor
  \State $R\gets L_1$.
  \For{$j=1,2,\ldots,d$}
    \If{$L_j=\varnothing$}
      \State \Return $(\texttt{cut},R)$
    \EndIf
    \If{$j$ is odd, $j+1\le d$, and $L_j\cap\mathcal T\ne\varnothing$}
      \State Choose $e\in L_j\cap\mathcal T$; $P\gets(\snk)$; $v\gets e$.
      \While{$v\ne\src$}
        \State Prepend $v$ to $P$; $v\gets\pi(v)$.
      \EndWhile
      \State Prepend $\src$ to $P$.
      \State \Return $(\texttt{path},P)$
    \EndIf
    \If{$j=d$}
      \State \Return $(\texttt{horizon},(L_1,\ldots,L_d))$
    \EndIf
    \State $L_{j+1}\gets\Call{NextLayer}{j+1,L_j}$.
  \EndFor
\EndFunction
\end{algorithmic}

\caption{BFS in the exchange graph up to distance $d$ using initial distance
lower bounds and recording the correct distances up to $d$}\label{alg:warm-bfs}
\end{algorithm}

If a completed layer is empty, no path can reach a later layer. Thus,
the returned $R$ contains all vertices reachable from $\src$ and
is a closed separating set. When layer $L_{d}$ is completed, the
algorithm has determined every vertex at distance at most $d$ and
makes no further search call.

\textbf{Retaining distance lower bounds across augmentations.} Algorithm~\ref{alg:alg:warm-sp}
is the complete \textsc{WarmSearch} called in Section~\ref{sec:near-nr-bound}.
At a path return, $R$ is exactly the union of the layers below the
shortest sink distance $\ell=|P|$, which is even. The algorithm caps
the distance lower bound of every undiscovered vertex at the smallest
integer at least $\ell$ of its \emph{old} membership parity: $\ell$
for vertices in $X$ and $\ell+1$ for vertices in $E\setminus X$.
It then augments and increases the distance lower bound of each path
vertex by one to account for its new parity (a path vertex that was
in $X$ before augmentation is now in $E\setminus X$ and vice versa).
The proof below shows that none of these operations decreases a distance
lower bound. The last loop explicitly refreshes the containing bases
in the underlying matroids at the unchanged prices; its oracle calls
are separate from the queries in Algorithm~\ref{alg:opt-query}.

\begin{algorithm}[tbh]
\begin{algorithmic}[1]
\Require $X\subseteq B_i\in\mathcal B(N_i(p_i))$; $r\ge1$, $d\ge2$, and $|X|\le\tau\le r$.
\Ensure Return $(\mathsf{status},X,B_1,B_2,\mathcal C)$ with the stated certificate.
\Function{WarmSearch}{$X,p,q,B_1,B_2,d,\tau$}
  \State $p_1\gets p$; $p_2\gets q$.
  \State $d_v\gets1$ for $v\notin X$ and $d_v\gets2$ for $v\in X$.
  \While{$|X|<\tau$}
    \State $(\mathsf{status},\mathcal C)\gets\Call{LayerSearch}{d}$.
    \If{$\mathsf{status}\ne\texttt{path}$}
      \State \Return $(\mathsf{status},X,B_1,B_2,\mathcal C)$
    \EndIf
    \State $P\gets\mathcal C$; $\ell\gets|P|$.
    \For{each $v\in E\setminus R$}
      \State $d_v\gets\ell$ if $v\in X$, and $d_v\gets\ell+1$ otherwise.
    \EndFor
    \State $X\gets X\triangle V_E(P)$.
    \For{each $v\in V_E(P)$}
      \State $d_v\gets d_v+1$.
    \EndFor
    \For{$i=1,2$}
      \State $B_i\gets X$.
      \For{each $e\in E\setminus X$, in nonincreasing order of $p_i(e)$}
        \If{$|B_i|<r$}
          \If{$\operatorname{InputInd}_i((B_i\cup\{e\})\setminus Z_{\rm dum})$}
            \State $B_i\gets B_i\cup\{e\}$.
          \EndIf
        \EndIf
      \EndFor
    \EndFor
  \EndWhile
  \State \Return $(\texttt{target},X,B_1,B_2,\varnothing)$
\EndFunction
\end{algorithmic}

\caption{Truncated BFS search in fixed price-optimal matroids}\label{alg:alg:warm-sp}
\end{algorithm}

\begin{lem}
[Truncated BFS search] \label{lem:truncated-warm-search} With the
stated inputs, Algorithm~\ref{alg:alg:warm-sp} repeatedly augments
along shortest paths until $\left|X\right|=\tau$, certifies a closed
separating set, or determines all distance layers through distance
$d$ with no source--sink path of length at most $d$. If $a$ augmentations
are performed, its total number of original independence queries,
including basis refreshes, is $O\bigl(n(a+d)\log(r+1)\bigr)$. 
\end{lem}

\begin{proof}
\emph{Exact layers and paths.} The scan for all arcs from $\src$
gives the exact first layer. Inductively suppose all layers below
$j$ are known. A vertex at distance $j$ must have an incoming arc
from $L_{j-1}$. For odd $j$, its valid lower bound is at most $j$,
so it is among the tested candidates. The first-matroid exchange test
decides exactly whether such a predecessor exists. Failure rules out
distance $j$; parity and the earlier completed layers then justify
increasing its label to $j+2$. For even $j$, the second-matroid
searches enumerate all arcs from $L_{j-1}$ to unreached inside vertices.
Each successful call removes one such vertex from $Q$, so every vertex
in the next layer is found exactly once. Its parent lies in the preceding
layer. This proves the layer induction and the validity of all lower-bound
updates during one search. Parent pointers give a shortest path to
every discovered vertex. The first reached neighbor of $\snk$ gives
the exact shortest $\src-\snk$ distance; following its parents therefore
produces a shortest, hence augmenting, path.

If $L_{j}$ is empty then no additional vertices can be reached. None
of the reached vertices are neighbors of $\snk$ so the returned $R$
is a closed separating set. When $L_{d}$ is completed, the algorithm
has obtained all layers through $d$ and can certify that there is
no source-sink path of length at most $d$.

\emph{Valid, nondecreasing distance lower bounds.} At a successful
return with sink distance $\ell$, every vertex of $R$ has its exact
distance below $\ell$, while every unreached vertex has distance
at least $\ell$. Setting the distance lower bounds for unreached
vertices to $\ell$ or $\ell+1$ according to the appropriate parity
is therefore valid before augmentation. Proposition~\ref{prop:import-distance}
says that every new distance is at least the minimum of its old distance
and $\ell$. An unchanged vertex keeps its parity, so its retained
exact or capped distance label remains valid. Each changed vertex
lies on $P$, had an exact distance below $\ell$, and switches parity.
Its new distance is therefore at least that old distance plus one,
justifying the increment in the algorithm.

It remains to check that capping never lowers a stored distance label.
For the first search the trivial lower bounds are at most two, which
is at most the $\src-\snk$ distance $\ell$. After a successful search
of length $\ell_{0}$, the labels on entering the next search are
at most $\ell_{0}$ inside and $\ell_{0}+1$ outside. This also holds
for path vertices after their parity increments. The next shortest
sink distance $\ell$ satisfies $\ell\ge\ell_{0}$, by the capped-distance
consequence proved in Section~\ref{sec:truncated-bfs}. Every discovered
vertex gets its label when a path to it is found so its label is at
most $\ell$. For an undiscovered outside vertex $e$, when the call
to \textsc{FindExchange} fails to find any arc from $L_{j-1}$ to
$e$, its label is increased to $j+2$. When the shortest path of
length $\ell$ is found, layer $L_{\ell-1}$ is the last layer constructed.
Thus, $j\le\ell-1$ and the new label $j+2\le\ell+1$. Every label
in the entire call is at most $d+2$: the algorithm only constructs
up to layer $L_{d}$ and the largest update due to a failed \textsc{FindExchange}
is $j+2\le d+2$, and a successful \textsc{FindExchange} sets the
label to $j\le d$.

\emph{Number of queries.} By Proposition~\ref{prop:import-neighbor},
one \textsc{FindExchange} call costs $O(\log(|T|+1))=O(\log(r+1))$
queries when $T$ is nonempty, and none when it is empty. Each internal
query costs at most one original query through \textsc{Probe}. In
one \textsc{LayerSearch}, scanning for arcs adjacent to $\src,\snk$
using \textsc{Probe} costs $O(n)$ queries. Successful odd-layer calls
to \textsc{FindExchange} discover distinct outside vertices, so there
are at most $n$ of them. Each unsuccessful odd-layer call to \textsc{FindExchange}
raises its tested vertex's distance label by two. Even-layer enumeration
has at most $|X|$ successful calls, since each discovers a distinct
inside vertex, and at most one terminating unsuccessful call per processed
outside vertex, hence at most $n$ more calls. Consequently one invocation
costs
\[
O\left(\left(n+\sum_{v\in E}\Delta d_{v}\right)\log(r+1)\right),
\]
where $\Delta d_{v}$ counts the actual distance label increases during
that invocation. There are at most $a+1$ invocations. Since all retained
labels are nondecreasing and bounded by $d+2$, their total increases,
including the caps and parity increments between invocations, sum
to $O(nd)$. The search cost is therefore $O(n(a+d)\log(r+1))$ queries.
The greedy extensions to compute optimal bases $B_{1},B_{2}\supseteq X$
use at most $2n$ queries per augmentation. Their $O(na)$ query cost
is a lower order term compared with the number of queries for search.
\end{proof}

\section{Finding a price update more efficiently}

\label{sec:thin-layer-repair}

In this section, we show that, by removing a small layer of the BFS
from $X$, we can perform a price update more efficiently. With appropriate
parameters, the cost from having to augment $X$ again in the future
is outweighed by the gain in the dual from the price update.

Suppose the current prices satisfy $p+q=c$, and let $N_{1}=N_{1}(p)$,
$N_{2}=N_{2}(q)$ be their respective matroids. The current common
independent set is $X$, with deficit $k=r-|X|$, and $B_{i}\supseteq X$
are bases in $N_{i}$.
\begin{lem}
[Small-layer repair] \label{lem:small-layer-repair} Suppose BFS
has found no source--sink path of length at most $2m$, and has determined
all BFS layers up to distance $2m$. For any $1\le j\le m$, let 
\[
L_{{\rm del}}=\{x\in X:\dist(\src,x)=2j\},\qquad A=\{e\in E:\dist(\src,e)\le2j\},\qquad Y=X\setminus L_{{\rm del}},
\]
and let $b=|L_{{\rm del}}|$. There is an integer $0\le b'\le b$
such that 
\[
\rk_{N_{1}}(E\setminus A)=|X\setminus A|+b',\qquad\rk_{N_{2}}(A)=|X\cap A|.
\]
One can choose a $p$-maximum basis $C_{1}\supseteq Y$ of $M_{1}$
such that $\left|C_{1}\setminus A\right|=\rk_{N_{1}}(E\setminus A)$,
using $O(n)$ original independence queries. The existing basis $B_{2}\supseteq X\supseteq Y$
already satisfies $\rk_{N_{2}}(A)=|X\cap A|=|B_{2}\cap A|$. After
replacing $X$ by $Y$, use the price update 
\[
p'=p-\mathbf{1}_{A},\qquad q'=q+\mathbf{1}_{A}.
\]
Then $C_{1},B_{2}$ are optimal for the new prices $p'$ and $q'$.
$Y$ is independent in both $N_{1}(p'),N_{2}(q')$, and the exact
dual decrease is 
\[
U(p,q)-U(p',q')=k-b'\ge k-b.
\]
An early exhaustive BFS gives the same conclusion with its set $A$
of all vertices reachable from $\src$ and $L_{{\rm del}}=\varnothing$. 
\end{lem}

\begin{proof}
Let $Q=(X\cap A)\setminus L_{{\rm del}}\subseteq X\cap A$ and $P=X\setminus Q=(X\setminus A)\cup L_{{\rm del}}$.

Every neighbor of $\src$ belongs to $A$, and $A$ contains no neighbor
of $\snk$. There is no arc from $(E\setminus X)\cap A$ to $X\setminus A$:
a vertex in $(E\setminus X)\cap A$ has distance at most $2j-1$,
so all its out-neighbors in $X$ have distance at most $2j$ and are
examined by the BFS. Likewise there is no arc from $Q=(X\cap A)\setminus L_{{\rm del}}$
to $\left(E\setminus X\right)\setminus A$, because all vertices in
$Q=(X\cap A)\setminus L_{{\rm del}}$ have distance from $\src$ at
most $2j-2$.

For $e\in A\setminus X$, the set $X+e$ is dependent in $N_{2}$,
because $e$ is not a neighbor of $\snk$. There is no arc from $e$
to $X\setminus A$ so its fundamental circuit in $X+e$ contains no
element of $X\setminus A$. Thus $(A\setminus X)\cup(X\cap A)=A\subseteq\cl_{N_{2}}(X\cap A)$,
proving the equality $\rk_{N_{2}}(A)=|X\cap A|$.

For $e\in(E\setminus A)\setminus X$, the set $X+e$ is dependent
in $N_{1}$, because $e$ is not a neighbor of $\src$. There is no
arc from $Q$ to $e$ so its fundamental circuit in $X+e$ contains
no element of $Q$. Thus, $((E\setminus A)\setminus X)\cup P=(E\setminus A)\cup L_{\mathrm{del}}\subseteq\cl_{N_{1}}(P)$.
Therefore, $\rk_{N_{1}}(E\setminus A)\le\left|P\right|=|X\setminus A|+b$.
Since $X\setminus A$ is independent, we have $\rk_{N_{1}}(E\setminus A)\ge\left|X\setminus A\right|$.
Let $\rk_{N_{1}}(E\setminus A)=|X\setminus A|+b'$ for some $b'$
satisfying $0\le b'\le b$.

Next we show that $Y$ is independent in $N_{1}$ and can be extended
to a basis $C_{1}$ in $N_{1}$. Since $P\cup Q=X$ is independent,
$Q$ is independent after contraction of $\cl_{N_{1}}(P)$. The inclusion
$E\setminus A\subseteq\cl_{N_{1}}(P)$ therefore implies 
\[
\rk_{N_{1}}((E\setminus A)\cup Q)=\rk_{N_{1}}(E\setminus A)+|Q|.
\]
Choose a basis $J$ of the restriction to $E\setminus A$ that contains
$X\setminus A$. Then $J\cup Q$ is independent, contains $Y$, and
extends to a basis $C_{1}$ in $N_{1}(p)$ with $\left|C_{1}\cap(E\setminus A)\right|=\rk_{N_{1}}\left(E\setminus A\right)$.

To compute $C_{1}$, run matroid greedy extension from the independent
set $Y$, sorting the remaining elements lexicographically by $(p(e),\mathbf{1}_{E\setminus A}(e))$
(i.e. greedy by price and breaking ties in favor of elements in $E\setminus A$).
This is greedy in the contraction $M_{1}/Y$ and costs at most $n$
original independence queries. Because $Y$ extends to a $p$-maximum
basis, the result is globally $p$-maximum. Among such bases containing
$Y$, it maximizes the number of elements outside $A$. The preceding
existence argument shows that this number is exactly $\rk_{N_{1}}(E\setminus A)$.
In particular $b'$ can be obtained from the resulting basis.

Next, we show $C_{1}$ is $p'$-maximum. $C_{1}$ is $p$-maximum
if and only if, for any valid exchange that removes $x$ and adds
$y$, the slack $p(x)-p(y)$ is nonnegative. The only exchange slacks
decreased by the update from $p$ to $p'$ remove an element of $A$
and insert an element outside $A$. A zero-slack exchange of this
type for $p$ would create a $p$-maximum basis with more than $\left|C_{1}\cap(E\setminus A)\right|=\rk_{N_{1}}(E\setminus A)$
elements outside $A$, which is impossible. Thus, the slack of every
valid exchange of this type in $p$ is at least one and the corresponding
slack in $p'$ is nonnegative. All other slacks stay nonnegative as
well. Thus $C_{1}$ is $p'$-maximum. The symmetric argument, using
$\left|B_{2}\cap A\right|=\rk_{N_{2}}(A)$, proves that $B_{2}$ is
$q'$-maximum. Both updated optimal bases contain $Y$, proving that
$Y$ is independent in both $N_{1}(p')$ and $N_{2}(q')$.

Evaluating the new optima on these same bases gives 
\begin{align*}
U(p',q')-U(p,q) & =-|C_{1}\cap A|+|B_{2}\cap A|\\
 & =-r+\rk_{N_{1}}(E\setminus A)+\rk_{N_{2}}(A)\\
 & =-r+|X|+b'=-k+b'.
\end{align*}
For a fully exhausted BFS, the same arguments apply with $L_{\mathrm{del}}=\varnothing$,
giving $b=b'=0$ and the closed separating set $A$ consisting of
all vertices reachable from $\src$.
\end{proof}

\section{Efficient implementation of the second stage of a scaling phase}

\label{sec:multicut-search}

Recall from Section \ref{subsec:overview} that a scaling phase is
comprised of two stages. In this section, we describe how to efficiently
implement the second stage when $r-\left|X\right|\leq\sqrt{r}$. In
this stage, the algorithm aims to find up to $\sqrt{r}$ augmenting
paths. This stage is comprised of epochs, where each epoch consists
of a sequence of price updates followed by a single augmentation.
In the following, we discuss how to efficiently implement each of
these epochs.

We start with some definitions. A \emph{fixed-price interval} is a
portion of an epoch in which the prices $p,q$ and thus the matroids
$N_{1}(p),N_{2}(q)$ remain unchanged. Each interval ends at a price
update or the final augmentation. During an \emph{epoch}, the common
independent set $X$ of $N_{1}(p),N_{2}(q)$ is fixed. A \emph{block}
is a portion of an epoch with a fixed heavy threshold and an allowance
of cuts/price updates. Throughout the epoch, the algorithm maintains
a region $F$ of vertices reachable from $\src$ and paths from $\src$
to all vertices in $F$ using only arcs inside $F$. Section \ref{subsec:tools}
describes the tools from prior work to deal with the single augmentation
and maintain $F$ within one fixed-price interval. Section \ref{sec:multicut-oracles}
shows that $F$ can be maintained through multiple price updates.
Section \ref{subsec:block-cost} bounds the cost of one block. Finally,
Section \ref{sec:epoch-analysis} uses the doubling trick to resolve
the issue that the number of price updates is unknown and bounds the
cost of an epoch.

\subsection{Tools from prior work}

\label{subsec:tools}

 The next proposition shows that one can find an augmenting path
if one exists efficiently regardless of the distance between $\src$
and $\snk$.
\begin{prop}
[fast single augmentation] \label{prop:import-fast} Given independence
oracles for two fixed matroids and a common independent set $X$,
there is an algorithm that can either find an augmenting path or correctly
certify that none exists with the following properties. For $\sigma:=\left|X\right|\ge1$,
the algorithm uses $O(n\sqrt{\sigma+1}\log\left(n+1\right))$ independence
queries with high probability \cite[Theorem 3.1]{bbmn}. It is always
correct, and the same bound holds in expectation \cite[Remark 3.9]{bbmn}.
Any returned simple path can be made chordless with $O((\sigma+1)\log(\sigma+1))$
further queries by the post-processing procedure in \cite[Section 3.1]{bbmn}.
\end{prop}

Recall from Section \ref{sec:exchange-graph} that, to perform a price
update, we need a closed separating set $A$ such that there is no
arc from $\{\src\}\cup A$ to $(E\setminus A)\cup\{\snk\}$ in the
exchange graph. We refer to $\{\src\}\cup A$ as a separating cut.
The next propositions define a growing set $F$ of vertices reached
from $\src$ maintained by the algorithm to either 1) obtain an augmenting
path; or 2) extend $F$ into a separating cut.  During a fixed-price
interval, the matroids $N_{1}(p),N_{2}(q)$ are fixed so the exchange
graph is fixed. Therefore, we can use the tools from \cite{bbmn},
which were developed for the unweighted case but are also applicable
here. In a fixed-price interval, the algorithm tries to expand $F$
(toward getting an augmenting path) and also tries to form a separating
cut (to get a price update). The categorization of vertices in $E\setminus X\setminus F$
into heavy and light below is used for the maintenance and expansion
of $F$ during a fixed-price interval.
\begin{defn}
\label{def:categorize}\cite[Definition 3.3]{bbmn} Fix an exchange
graph for a fixed $X$, a threshold $h\ge1$, and a growing set $F$.
A vertex $e\notin X\cup F$ is called \emph{heavy} if either $(e,\snk)$
is an arc of the graph or $e$ has at least $h$ out-neighbors in
$X\setminus F$. Otherwise, $e$ is called \emph{light}.
\end{defn}

\begin{prop}
[reverse BFS to expand reach or produce cut] \label{prop:import-reverse}
Let $F\subseteq E\cup\{\src\}$ be a set of vertices that are reachable
from $\src$ in the exchange graph. Let $F_{X}=F\cap X$ and $F_{{\rm out}}=F\cap(E\setminus X)$.
Suppose no arc goes from $F_{{\rm out}}$ to $X\setminus F_{X}$.
Partition the vertices in $E\setminus(X\cup F)$ into heavy and light
vertices, and suppose all outgoing arcs of every light vertex into
$X\setminus F_{X}$ are known. A reverse BFS from all heavy vertices
\cite[Algorithm 3.7]{bbmn} finds a path from $F$ to a heavy vertex,
or certifies there is none, using $O(n(1+\log(\sigma+1)))$ independence
queries where $\sigma=\left|X\right|$ \cite[Lemma 3.6]{bbmn}. 
\end{prop}

The search algorithm \cite[Algorithm 3.8]{bbmn} also 1) keeps $\snk\notin F$;
2) adds all out-neighbors of $F_{{\rm out}}$ to $F$, stopping with
an augmenting path if any vertex in $F_{\mathrm{out}}$ has an arc
to $\snk$; and 3) treats every sink-neighbor as heavy. These additional
invariants turn an unsuccessful reverse BFS into a closed separating
set. They are stated explicitly in the following lemma.
\begin{lem}
[Reverse BFS certificate]\label{lem:reverse} Suppose a failed search
maintains a reached set $F$ containing $\src$ and excluding $\snk$,
with a recorded path inside $F$ from $\src$ to each vertex in $F$,
with no arc from $F\setminus X$ to $(X\cup\{\snk\})\setminus F$.
Partition the vertices in $E\setminus(X\cup F)$ into heavy and light
vertices, where all neighbors of $\snk$ are heavy and every light
vertex's outgoing arcs to $X\setminus F$ are known. Run reverse BFS
from all heavy vertices, enumerating predecessors of vertices outside
$X$ and using the known light arcs into vertices inside $X$. If
this search finishes without meeting $F$, its visited set $R_{{\rm rev}}$
makes $A=E\setminus R_{{\rm rev}}$ a closed separating set. 
\end{lem}

\begin{proof}
Every in-neighbor of a visited vertex outside $X$ is visited by enumeration
of the reverse BFS. An unvisited in-neighbor of a visited vertex inside
$X$ cannot be heavy, because all heavy vertices are roots of the
reverse BFS. It cannot belong to $F$, by the assumed invariant. Thus
it is light and is visited through its known arc. Thus, all in-neighbors
of $R_{{\rm rev}}$ are in $R_{{\rm rev}}$. It avoids $\src\in F$,
so it avoids all neighbors of $\src$, and it contains all neighbors
of $\snk$ as roots. Thus, $A=E\setminus R_{{\rm rev}}$ is a closed
separating set.
\end{proof}

The final unsuccessful reverse BFS in Algorithms 3.7--3.8 of~\cite{bbmn}
meets this lemma's hypotheses. The algorithm uses sampling to estimate
which vertices are heavy, but it ensures that the set $A$ is a closed
separating set with probability one as follows: the algorithm explicitly
marks all neighbors of $\snk$ as heavy and enumerates all of the
out-neighbors of a vertex that is marked as light.

The next proposition bounds the cost of categorizing the vertices
outside $F$ and the amortized cost of enumerating all out-neighbors
of light vertices as $F$ grows in one fixed-price interval.
\begin{prop}
[heavy/light categorization cost] \label{prop:import-categorize}
\cite[Lemma 3.4 and Section 5.1]{bbmn} Fix an exchange graph for
a fixed set $X$, a threshold $h\ge1$, and a growing set $F$. For
a vertex $e\notin(X\cup F)$, let 
\[
H_{e}=\bigl|\{x\in X\setminus F:e\to x\text{ is an arc}\}\bigr|.
\]
After explicitly categorizing the vertices with arcs to $\snk$ as
heavy, the sampling routine uses $O(n\log(n+1))$ queries per categorization
phase. For any prescribed constant $C>0$, its sampling constants
can be chosen so that, except with probability at most $(n+1)^{-C}$
over the polynomially many decisions, both of the following hold:
\begin{enumerate}
\item every vertex marked as heavy via sampling has $H_{e}\ge h$;
\item every vertex whose full neighborhood is enumerated has $H_{e}\le10h$
at that enumeration.
\end{enumerate}
Vertices not marked heavy by sampling have their full out-neighbor
lists enumerated (or use an already known list), and are then classified
by their exact current degree: heavy for $H_{e}\ge h$, light for
$H_{e}<h$. The total enumeration cost while $F$ grows in this fixed
graph is $O(nh\log(n+1))$ on the good event. The out-neighbor list
of a vertex marked as light is complete for every sampling outcome,
including bad outcomes.
\end{prop}

\cite[Experiment 5.1 and Claim 5.2]{bbmn} justify the two degree
guarantees. For completeness, we include the sampling argument here.
Let $u=|X\setminus F|$. If $u>10h$, draw 
\[
q_{{\rm samp}}=\left\lceil \frac{\log(1/4)}{\log(1-10h/u)}\right\rceil 
\]
elements uniformly and independently with replacement from $X\setminus F$,
and let $R$ be the set of distinct elements drawn. The probability
that $R$ avoids the outgoing neighbors of $e$ is $(1-H_{e}/u)^{q_{{\rm samp}}}$.
This probability is at least $3/4$ when $H_{e}<h$, and at most $1/4$
when $H_{e}>10h$; $L=O(\log(n+1))$ independent repetitions separate
these cases. Testing if $e$ has a neighbor in $R$ is implemented
by checking independence of $(X\setminus R)+e$ for a sampled subset
$R\subseteq X\setminus F$. For a vertex $e$ not neighboring $\snk$,
dependence of this set means that $R$ avoids its outgoing neighbors.
If $|X\setminus F|\le10h$, we can directly enumerate the neighbors
of $e$ in $X\setminus F$ without sampling.

Cached complete out-neighbor lists can be reused as $F$ grows because
the candidate neighborhoods $X\setminus F$ only shrink. They must
be rebuilt after a price change since new arcs might appear due to
the price change.

\subsection{Maintenance of $F$ through price updates}

\label{sec:multicut-oracles}

The set $F\subseteq E\cup\{\src\}$ maintained by \cite[Algorithms 3.7–3.8]{bbmn}
contains $\src$ and comes with recorded paths from $\src$ to every
vertex in $F$ using only arcs inside $F$. In this section, we show
that these paths remain valid after a price update.

Before a reverse-search phase, the set $F$ satisfies
\begin{equation}
\text{there is no arc from }F\setminus(X\cup\{\src\})\text{ to }(X\cup\{\snk\})\setminus F.\label{eq:multicut-F-closure}
\end{equation}
As discussed above, a reverse search from heavy vertices that does
not reach $F$ supplies the closed separating set $A=E\setminus R_{{\rm rev}}$,
where $R_{\mathrm{rev}}$ is the set of vertices reached by the reverse
search. In particular, $F\setminus\{\src\}\subseteq A$. The price
update is $p'\gets p-\ind_{A}$, $q'\gets q+\ind_{A}$.
\begin{lem}
[Recorded paths remain valid] \label{lem:multicut-path-retention}
Let $A\subseteq E$ be a closed separating set, and let $p'=p-\ind_{A}$,
$q'=q+\ind_{A}$ be the updated prices. Every old exchange arc whose
endpoints lie on the same side of $A$ remains valid in the new matroids
$N_{1}(p'),N_{2}(q')$. Every old arc adjacent to $\src$ or $\snk$
also remains valid. Consequently the recorded paths from $\src$ to
all vertices in $F$ remain valid after the price update. 
\end{lem}

\begin{proof}
As shown in Lemma \ref{lem:closed}, the closed separating set is
a min--max witness, and thus
\[
|X\setminus A|=\rk_{N_{1}(p)}(E\setminus A),\qquad|X\cap A|=\rk_{N_{2}(q)}(A).
\]
A basis is $p$-maximum if and only if every feasible single exchange
removing $x$ and inserting $y$ has nonnegative slack $p(x)-p(y)$. 

A $p$-maximum basis $C$ that satisfies $\left|C\setminus A\right|=|X\setminus A|=\rk_{N_{1}(p)}(E\setminus A)$
remains optimal for $p'$. Indeed, the only exchange slacks of $C$
that decrease remove an element $x$ in $A$ and insert an element
$y$ outside $A$. A zero-slack exchange of this type would be another
$p$-maximum basis with more elements outside $A$ than $\rk_{N_{1}(p)}(E\setminus A)$,
which is a contradiction. Every such slack for $p$ is therefore at
least one, and the unit update makes the corresponding slack in $p'$
nonnegative. The corresponding statement for $q\gets q+\ind_{A}$
uses optimal bases $C$ with $|C\cap A|=|X\cap A|=\rk_{N_{2}(q)}(A)$.

For an old exchange arc with endpoints $x\in X$ and $e\notin X$
on the same side of $A$, let $Y=X-x+e$. Since $x$ and $e$ lie
on the same side, $|Y\setminus A|=|X\setminus A|$ and $|Y\cap A|=|X\cap A|$.
If the arc is $x\to e$, then $Y$ is independent in $N_{1}(p)$.
Choose a basis $C$ of $N_{1}(p)$ containing $Y$. We have
\[
|X\setminus A|=|Y\setminus A|\le|C\setminus A|\le\rk_{N_{1}(p)}(E\setminus A)=|X\setminus A|.
\]
Thus $C$ saturates $E\setminus A$ in $N_{1}(p)$ and remains $p'$-maximum
by the preceding argument. Thus $Y$ is independent in $N_{1}(p')$
and hence $x\to e$ remains an arc in the exchange graph.

If the arc is $e\to x$, then $Y$ is independent in $N_{2}(q)$.
Choose a basis $C$ of $N_{2}(q)$ containing $Y$. We have
\[
|X\cap A|=|Y\cap A|\le|C\cap A|\le\rk_{N_{2}(q)}(A)=|X\cap A|.
\]
Thus $C$ saturates $A$ in $N_{2}(q)$ and remains $q'$-maximum.
Thus $Y$ is independent in $N_{2}(q')$ and hence $e\to x$ remains
an arc in the exchange graph.

For an old arc $\src\to e$, the cut has $e\in A$, and $X+e$ is
$N_{1}(p)$ independent. We have $\left|(X+e)\setminus A\right|=\left|X\setminus A\right|$
so any $p$-maximum basis $C$ containing $X+e$ remains optimal for
$p'$. Thus, the arc $\src\to e$ remains valid.

For an old arc $e\to\snk$, we have $e\notin A$. We have $\left|\left(X+e\right)\cap A\right|=\left|X\cap A\right|$
so any $q$-maximum basis $C$ containing $X+e$ remains optimal for
$q'$. Thus, the arc $e\to\snk$ remains valid. 

Every edge of a recorded source path inside $F$ is covered by these
conclusions. 
\end{proof}

After a price update, we keep $F$ and its recorded paths, discard
all cached neighbor lists, and restore (\ref{eq:multicut-F-closure})
as follows. For each vertex outside $X$ that is already in $F$,
use the $\textsc{OutEdge}$ subroutine from Proposition \ref{prop:import-neighbor}
to enumerate its current neighbors in $(X\cup\{\snk\})\setminus F$,
adding each newly found vertex inside $X$ to $F$ with its witnessing
edge. Discovery of $\snk$ immediately gives an augmenting walk. There
are at most $n$ unsuccessful neighbor searches per price update,
and every vertex inside $X$ is added at most once while $F$ is maintained.
Hence, over $D$ price updates, restoration costs $O\bigl((n(D+1)+r)\log(r+1)\bigr)$
queries. During a successful reverse-search phase, the same restoration
operation is applied to the newly reached vertices outside $X$; every
vertex outside $X$ is new only once. Its additional cost is $O((n+r)\log(r+1))$
queries. Recorded paths yield an augmenting walk when $\snk$ is found;
removing cycles from the walk and converting it into a chordless path
cost $O(r\log(r+1))$ queries \cite[Section 3.1]{bbmn} and occur
only at the end of the epoch.

\subsection{Analysis of the number of queries in a block}

\label{subsec:block-cost}

Consider a block allowing at most $D$ price updates, with $0\le D\le r+1$.
It starts with $F=\{\src\}$ and a fresh categorization state. Alternatively,
it can start with any $F$ with paths inside $F$ from $\src$ to
all of $F$ and satisfying (\ref{eq:multicut-F-closure}). Fix an
integer $1\le h\le r$. We use the heavy/light categorization of Proposition~\ref{prop:import-categorize}
with this value $h$. Recall that every neighbor of $\snk$ is heavy.
Additionally, on the good sampling event, all other heavy vertices
have at least $h$ out-neighbors in $X\setminus F$, and light vertices
have fewer than $h$. A vertex not marked heavy by sampling may nevertheless
be marked heavy after its neighbor list is enumerated during categorization;
the good event bounds every such enumerated list by $10h$. On every
outcome, the neighbor lists of all vertices marked as light are complete.
The block stops at an augmentation or immediately after its $D$th
price update. Neighbor lists of light vertices are discarded after
each price update, but $F$ is maintained throughout the block.

A successful reverse-search phase finds a path from $F$ to a heavy
vertex and adds to $F$ that path and all the current out-neighbors
of every vertex on the path that is outside $X$. If the heavy vertex
has an arc to $\snk$, the epoch ends. Otherwise a correctly categorized
heavy vertex adds at least $h$ previously unreached vertices that
are inside $X$. Thus there are at most $r/h$ such successful phases
in the entire block. There are at most $D+1$ reverse-search phases
that fail to reach $F$ or find an augmenting walk.

Within a single fixed-price interval, the set $(X\cup\{\snk\})\setminus F$
shrinks as $F$ grows. Once the entire list of current out-neighbors
of a vertex outside $X$ is enumerated, that list remains usable for
the rest of this interval. The list is discarded at the next price
update. This is the only rebuilding charged once per fixed-price interval.

Putting everything together, we obtain the following upper bound on
the number of queries in the block.
\begin{lem}
[Number of queries in a block] \label{lem:multicut-random-block}
Consider a block. Conditioned on the state before the block, the expected
number of original independence queries performed during the block
is 
\[
O\left(n\log(n+1)\left(\frac{r}{h}+h(D+1)+D+1\right)\right).
\]
The block is always correct; randomization affects only its query
count. 
\end{lem}

\begin{proof}
Use the sampled heavy/light categorization from Definition \ref{def:categorize}
and Proposition~\ref{prop:import-categorize} with fresh samples
and a sufficiently large constant in its $O(\log(n+1))$ repetitions.
Neighbors of $\snk$ are identified explicitly and always treated
as heavy. On the event that all categorization guarantees hold, every
sampled heavy vertex without an arc to $\snk$ has at least $h$ neighbors,
and every vertex whose full out-neighbor list is enumerated has at
most $10h$ current neighbors.

There are at most $r/h+D+1$ reverse-search phases on this event.
Each costs $O(n\log(r+1))$ queries for reverse BFS and $O(n\log(n+1))$
queries for sampling. Complete neighbor enumeration costs $O(nh(D+1)\log(r+1))$
queries, because a vertex's full out-neighbor list need be enumerated
at most once in each of the $D+1$ fixed-price intervals. The restoration
cost is as established above. Adding up these costs gives the bound
in the lemma.

Next we argue that the algorithm is always correct and the query bound
holds in expectation in addition to with high probability. Regardless
of sampling outcomes, a successful reverse search reaching $F$ but
not finding an augmenting walk adds at least one new outside vertex
to $F$: the starting vertex of the search was outside $F$. Hence
there are at most $n$ successful phases, and at most $D+1$ other
phases. With $D\le r+1\le n+1$, there are $O(n^{2})$ classification
decisions, each having its conditional error probability bounded by
an arbitrarily large inverse polynomial in $n+1$. Take the constants
large enough that a union bound gives total bad-event probability
at most $(n+1)^{-6}$.

Even on a bad sampling event, every full enumeration has at most $r$
neighbors, every such list is cached for its fixed-price interval,
and the number of phases is at most $O(n+D+1)$. Thus the worst query
count is $O(n^{3}\log(n+1))$, uniformly over all sampling outcomes.
Multiplying this by $(n+1)^{-6}$ gives an $O(1)$ bad-event contribution
to expectation, absorbed by the stated bound in the lemma. All probabilities
are conditional on the block's entry state. Correctness needs only
that vertices marked as light have their actual outgoing lists enumerated
and that every neighbor of $\snk$ is a root of the reverse search;
these facts hold without the sampling guarantees.
\end{proof}

\subsection{Analysis of the number of queries in an epoch}

\label{sec:epoch-analysis}

In this section, we show how to select the parameters $D$ and $h$
for each block of an epoch, and upper bound the expected number of
queries in the epoch.

Let $t$ denote the total number of price updates performed during
the entire epoch before its terminating augmentation. By Lemma \ref{lem:unit}
and $\Gamma\leq r$, we have $t\le r$.

To optimize the query bound in Lemma \ref{lem:multicut-random-block},
we need to set the heavy threshold $h$ as a function of the number
of price updates $D$, which is not known a priori. We use the doubling
trick to set $h$ and $D$ as follows.

Index the blocks by $j=0,1,2,\ldots$ and use price-update allowances
$D_{j}=2^{j}$. The allowance counts updates made \emph{within that
block}. A block ends either at the first augmentation, which ends
the epoch, or immediately after making its $D_{j}$th price update.
In the latter case, start a fresh block with allowance $D_{j+1}=2D_{j}$.
At the beginning of each block, reset $F=\{\src\}$. For a block with
allowance $D$, choose 
\[
h=\max\{1,\lceil\sqrt{r/(D+1)}\rceil\}.
\]
Let $g(D)=\sqrt{r(D+1)}+D+1$, which will be used to bound the cost
of the block. The inequalities $\sqrt{r/(D+1)}\le h\le\sqrt{r/(D+1)}+1$
give 
\[
\frac{r}{h}+h(D+1)+D+1\le2\sqrt{r(D+1)}+2(D+1)=2g(D).
\]
Thus Lemma~\ref{lem:multicut-random-block} bounds the expected block
cost by $O(n\log(n+1)g(D))$, conditional on the state before that
block.
\begin{claim}
\label{claim:multicut-envelope} Fix any realized execution of an
epoch, and let $J$ be the index of its last executed block, and $t$
be the number of price updates in the epoch. We have $D_{J}\le t+1$
and $\sum^{J}_{j=0}g(D_{j})=O(\sqrt{r(t+1)})$.
\end{claim}

\begin{proof}
Every block before block $J$ must have exhausted its allowance so
the number of price updates made \emph{before} block $J$ starts is
exactly $\sum^{J-1}_{j=0}D_{j}=1+2+\cdots+2^{J-1}=2^{J}-1=D_{J}-1$,
where the sum is zero when $J=0$. Let $u$ be the number of price
updates in the last block before its augmentation. The immediate stopping
rule gives $0\le u<D_{J}$: if it made $D_{J}$ updates, that block
would end and the augmentation would have to occur in a subsequent
block. Hence 
\[
t=D_{J}-1+u,\qquad D_{J}\le t+1<2D_{J}.
\]
In particular, $t\le r$ implies $D_{J}\le r+1$, as required by Lemma
\ref{lem:multicut-random-block}.

Next we bound $\sum^{J}_{j=0}g(D_{j})$. Since $D_{j}+1\le2\cdot2^{j}$,
the two parts of $g(D_{j})$ satisfy 
\begin{align*}
\sum^{J}_{j=0}\sqrt{r(D_{j}+1)} & \le\sqrt{2r}\sum^{J}_{j=0}2^{j/2}=O(\sqrt{rD_{J}}),\\
\sum^{J}_{j=0}(D_{j}+1) & \le2\sum^{J}_{j=0}2^{j}=O(D_{J}).
\end{align*}
Consequently,
\[
\sum^{J}_{j=0}g(D_{j})=O(\sqrt{rD_{J}}+D_{J})=O(\sqrt{r(t+1)}+t+1)=O(\sqrt{r(t+1)}).
\]
The last step uses $t+1\le r+1\le2r$, which implies $t+1\le\sqrt{2}\sqrt{r(t+1)}$.
\end{proof}

Finally, we upper bound the expected cost of an epoch.
\begin{prop}
[Expected cost of an epoch] \label{prop:multicut-epoch} Consider
an epoch and let $\mathcal{H}_{{\rm entry}}$ be the information at
its start. Let $t\le r$ be the number of price updates in the epoch.
Let $Q_{{\rm epoch}}$ be the number of queries made in the epoch.
For a universal constant $C'$, we have
\[
\mathbb{E}[Q_{{\rm epoch}}\mid\mathcal{H}_{{\rm entry}}]\le C'n\log(n+1)\,\mathbb{E}\!\left[\sqrt{r(t+1)}\,\middle|\,\mathcal{H}_{{\rm entry}}\right].
\]
The algorithm is always correct. The $O(n)$ queries to update the
bases $B_{1},B_{2}\supseteq X$ after augmentation are absorbed in
this bound.
\end{prop}

\begin{proof}
 There are only $1+\lfloor\log_{2}(r+1)\rfloor$ possible block indices,
since an executed block satisfies $D_{j}\le r+1$. For bookkeeping,
imagine assigning an independent random tape $\xi_{j}$ to each possible
block; a tape is the sequence of random choices the block will use.
These choices can still be generated on demand. Define 
\[
\mathcal{H}_{j}=\sigma(\mathcal{H}_{{\rm entry}},\xi_{0},\ldots,\xi_{j-1}),
\]
where $\sigma(\cdot)$ means the information generated by the listed
history and random choices. Let $I_{j}\in\{0,1\}$ indicate whether
block $j$ is executed, and let $Q_{j}$ be its query count, set to
zero when $I_{j}=0$. Whether block $j$ is reached, and its starting
state if it is reached, are determined by $\mathcal{H}_{j}$. Its
own tape $\xi_{j}$ is fresh and independent of that history.

If $I_{j}=0$ then $\mathbb{E}[Q_{j}\mid\mathcal{H}_{j},I_{j}=0]=0$.
If $I_{j}=1$ then as argued above using Lemma \ref{lem:multicut-random-block},
$\mathbb{E}[Q_{j}\mid\mathcal{H}_{j},I_{j}=1]\le Cn\log(n+1)\,g(D_{j})$
for a universal constant $C$. Thus, in both cases,
\[
\mathbb{E}[Q_{j}\mid\mathcal{H}_{j}]\le I_{j}\,Cn\log(n+1)\,g(D_{j}).
\]
Note that $Q_{{\rm epoch}}=\sum_{j}Q_{j}$. Since $\mathcal{H}_{{\rm entry}}\subseteq\mathcal{H}_{j}$,
the law of total expectation (average first over a block's fresh randomness,
then over its possible starting histories) gives 
\begin{align*}
\mathbb{E}[Q_{{\rm epoch}}\mid\mathcal{H}_{{\rm entry}}] & =\sum_{j}\mathbb{E}\!\left[\mathbb{E}[Q_{j}\mid\mathcal{H}_{j}]\,\middle|\,\mathcal{H}_{{\rm entry}}\right]\\
 & \le Cn\log(n+1)\,\mathbb{E}\!\left[\sum_{j}I_{j}g(D_{j})\,\middle|\,\mathcal{H}_{{\rm entry}}\right]\\
 & \le C'n\log(n+1)\,\mathbb{E}\!\left[\sqrt{r(t+1)}\,\middle|\,\mathcal{H}_{{\rm entry}}\right].
\end{align*}
Here $C'$ is another absolute constant, and the last inequality applies
Claim \ref{claim:multicut-envelope}. 
\end{proof}

\section{Algorithm for a single scaling phase}

\label{sec:near-nr-bound}

In this section, we describe the algorithm for one scaling phase using
$O\bigl(nr\log(n+1)\sqrt{\log(r+1)}\bigr)$ independence queries in
expectation.

\subsection{The two stages of one scaling phase}

Let $\beta\in\{0,1\}^{E}$ be the next revealed bit of the weights
of all items. Let $p_{0}=2p$, $q_{0}=2q+\beta$ and let $B_{0}$
be the optimal common basis for $p,q$ from the previous phase or
initialization. Keep $B_{1}=B_{0}$, choose $B_{2}$ greedily for
$q_{0}$ with ties favoring elements in $B_{0}$, and let $X_{0}=B_{0}\cap B_{2}$.
Let
\[
\Gamma=U(p_{0},q_{0})-c'(B_{0})=r-|X_{0}|\le r.
\]
If $\Gamma=0$, the reference basis already certifies the new optimum.
Choose $1\le K\le r$, and let 
\[
m=\left\lceil \frac{2r}{K}\right\rceil ,d=2m.
\]
The algorithm uses the following stages in order.
\begin{enumerate}
\item \emph{High deficit.} While $k=r-|X|>K$, run the truncated BFS algorithm
from Section \ref{sec:truncated-bfs} in the fixed current price-optimal
matroids, up to maximum distance $d$. Augment along its shortest
paths, refreshing the containing optimal bases after each augmentation.
Stop this stage immediately on reaching $k\le K$. If a closed separating
set is returned first, make the price update and restart the BFS routine.
If the search reaches the maximum distance $d$ first, choose a smallest
even layer among the first $m$ even layers. Apply the small-layer
repair procedure from Lemma \ref{lem:small-layer-repair}: remove
that layer from $X$, update the containing bases, make the unit price
update, and restart the BFS routine.
\item \emph{Low deficit.} This stage applies once $k\le K$. Use the algorithm
from Section \ref{sec:multicut-search} to efficiently implement each
epoch until $|X|=r$.  Each epoch maintains a partial set $F$ of
reachable vertices and paths inside $F$ from $\src$ to all of $F$
across a sequence of price updates and ends at its next augmentation. 
\end{enumerate}
When the truncated BFS reaches the maximum distance $d$, all vertices
at distance at most $d$ have been identified, and there is no $\src-\snk$
path of length at most $d$. The chosen even layer has at most $r/m\le K/2$
vertices. Here $K$ is the cutoff between high and low deficits, $m$
is the number of even layers examined, and $d=2m$ is the search horizon
measured in arcs. An empty layer is allowed and gives an ordinary
closed separating set with no deletion. When an empty layer is found,
BFS can stop immediately. When a path to $\snk$ is found, BFS first
completes the last layer and also stops.

\textbf{Pseudocode and subroutine guarantees.} Algorithms~\ref{alg:refine-bit}--\ref{alg:multicut-epoch}
spell out the implementation with $K=\lceil\sqrt{r}\rceil$. After
padding the matroids with dummy elements, $E,M_{1},M_{2},r$ are shared
by all procedures. \textsc{GreedyBasis} and \textsc{GreedyExtend}
use the underlying matroid independence oracles; the latter starts
from the specified independent set and scans the other elements in
the stated descending order. A \emph{basis refresh} means extending
the current $X$ greedily to a $p$-maximum basis $B_{1}$ and a $q$-maximum
basis $B_{2}$. All auxiliary neighborhood and terminal tests use
Lemma~\ref{lem:one-query}, as discussed in Section~\ref{sec:multicut-oracles}.

The call \textsc{WarmSearch} in Algorithm~\ref{alg:refine-bit} is
the routine of Lemma~\ref{lem:truncated-warm-search}, with target
$|X|\ge r-K$. It performs all intervening shortest-path augmentations
and basis refreshes, retaining the capped distance labels across those
augmentations. It returns $X,B_{1},B_{2}$, a certificate $\mathcal{C}$,
and a status: \texttt{target} if the target cardinality is reached;
\texttt{cut} with a set $A\subseteq E$ of all vertices reachable
from $\src$; or \texttt{horizon} with the layers $L_{i}=\{e\in E:\operatorname{dist}(\src,e)=i\}$
for $1\le i\le d$ and a certificate that the distance to $\snk$
exceeds $d$. Its distance labels reset only after a price update.

\begin{algorithm}[tbh]
\begin{algorithmic}[1]
\Require A common basis $B_0$ optimal for each of $p,q$, and the next bit $\beta$.
\Ensure A common basis optimal for the new weight vector $2(p+q)+\beta$,
         with its exact split.
\Function{RefineBit}{$B_0,p,q,\beta$}
  \State $p\gets2p$; $q\gets2q+\beta$; $B_1\gets B_0$.
  \State $B_2\gets\Call{GreedyBasis}{M_2,q}$, breaking equal-price ties in favor of $B_0$.
  \State $X\gets B_0\cap B_2$; $\Gamma\gets r-|X|$.
  \If{$\Gamma=0$}
    \State \Return $(X,p,q)$
  \EndIf
  \State $K\gets\lceil\sqrt r\rceil$; $m\gets\lceil2r/K\rceil$; $d\gets2m$.
  \While{$r-|X|>K$}
    \State $(\mathsf{status},X,B_1,B_2,\mathcal C)
       \gets\Call{WarmSearch}{X,p,q,B_1,B_2,d,r-K}$.
    \If{$\mathsf{status}=\texttt{target}$}
      \State \textbf{break}
    \ElsIf{$\mathsf{status}=\texttt{cut}$}
      \State $A\gets$ the closed separating set in $\mathcal C$.
    \Else
      \Comment{Horizon certificate; layers are for the current $X$}
      \State Read $L_1,\ldots,L_d$ from $\mathcal C$.
      \State Choose $j\in\arg\min_{1\le a\le m}|L_{2a}|$;
             $A\gets\bigcup_{i=1}^{2j}L_i$.
      \State $Y\gets X\setminus L_{2j}$.
      \State $B_1\gets\Call{GreedyExtend}{M_1,Y,(p,\ind_{E\setminus A})}$.
      \Comment{Old prices; breaking ties in favor of $E\setminus A$}
      \State $X\gets Y$; retain $B_2$.
      \Comment{Lemma~\ref{lem:small-layer-repair}}
    \EndIf
    \State $p\gets p-\ind_A$; $q\gets q+\ind_A$.
    \Comment{The stored bases remain optimal}
    \State Discard the old distance state before the next \textsc{WarmSearch} call.
  \EndWhile
  \While{$|X|<r$}
    \State $(X,p,q,B_1,B_2)\gets\Call{MulticutEpoch}{X,p,q,B_1,B_2}$.
  \EndWhile
  \State \Return $(X,p,q)$
\EndFunction
\end{algorithmic}

\caption{One bit phase: high-deficit repairs, then low-deficit epochs}\label{alg:refine-bit}
\end{algorithm}

Algorithm~\ref{alg:multicut-epoch} describes the implementation
of one epoch, including the doubling of price update allowances of
Section~\ref{sec:epoch-analysis}. The retained state $F$ includes
paths from $\src$ to all vertices in $F$. ``Resume the retained
search'' means the procedure of Section~\ref{sec:multicut-search}:
restore the closure condition (\ref{eq:multicut-F-closure}), enumerate
complete lists for vertices marked as light, and perform the heavy/light
reverse-search phases until an $\src-\snk$ walk or a closed separating
set is found. Every neighbor of $\snk$ is treated as heavy. New categorizations
draw fresh samples; a new block starts with randomness independent
of the prior history. Finding an $\src-\snk$ walk stops the search
immediately, just as finding one in a reverse-search phase does.

\begin{algorithm}[tbh]
\begin{algorithmic}[1]
\Require $|X|<r$ and optimal bases $B_1,B_2\supseteq X$ for the current prices $p,q$.
\Ensure One augmentation of $X$, with updated prices and containing optimal bases.
\Function{MulticutEpoch}{$X,p,q,B_1,B_2$}
  \State $D\gets1$.
  \Loop
    \State $h\gets\max\{1,\lceil\sqrt{r/(D+1)}\rceil\}$; $u\gets0$.
    \State $F\gets\{\src\}$; discard old search data and start a fresh randomized block.
    \While{$u<D$}
      \State Resume the retained search from $F$ with threshold $h$ until a walk or cut is found.
      \If{a source--sink walk is found}
        \State Remove cycles and convert the walk to a chordless augmenting path $P$.
        \State $X\gets X\mathbin{\triangle}(V(P)\cap E)$.
        \State Refresh $B_1,B_2$ at the current prices to contain the new $X$.
        \State \Return $(X,p,q,B_1,B_2)$
      \Else
        \State $A\gets E\setminus R_{\rm rev}$, the certified closed separating set.
        \State $p\gets p-\ind_A$; $q\gets q+\ind_A$; $u\gets u+1$.
        \State Retain $F$ and its source paths; keep $X,B_1,B_2$ unchanged.
        \State Discard all cached neighbor lists; rebuild them if this block continues.
      \EndIf
    \EndWhile
    \State $D\gets2D$.
    \Comment{Start the next block with twice the previous allowance}
  \EndLoop
\EndFunction
\end{algorithmic}

\caption{One low-deficit epoch with retained paths and doubling cut allowances}\label{alg:multicut-epoch}
\end{algorithm}

Here $u$ counts price updates in the current block, $R_{{\rm rev}}$
is the reverse-reachable set from the failed reverse search of Lemma~\ref{lem:reverse},
$V(P)$ is the vertex set of the extracted path, and $\triangle$
denotes symmetric difference. Each call to \textsc{MulticutEpoch}
resets $D$ to one and increases $|X|$ by exactly one. In particular,
the search state is retained across price updates within a block,
while the allowance is reset only at an epoch boundary; no deletion
occurs in this procedure.

\subsection{Analysis of the number of queries in the high-deficit stage}

In this section, we upper bound the number of queries in the high-deficit
stage. For repair $j$, let $k_{j}$ be the deficit just before deletion,
$b_{j}$ the number of elements removed from $X$, and $\delta_{j}=\rk_{N_{1}(p)}(E\setminus A)-|X\setminus A|$.
Lemma \ref{lem:small-layer-repair} gives 
\[
0\le\delta_{j}\le b_{j}\le K/2,\qquad U_{{\rm old}}-U_{{\rm new}}=k_{j}-\delta_{j}.
\]
Treat an ordinary separating update as $b_{j}=\delta_{j}=0$. Because
every high-stage update has $k_{j}>K$, we have
\[
k_{j}-\delta_{j}\ge k_{j}-b_{j}>K/2,\qquad k_{j}-\delta_{j}\ge b_{j}.
\]
The starting basis $B_{0}$ remains feasible throughout the bit scaling
phase and thus we have $U\ge c'(B_{0})$ throughout. For $\Gamma$
being the initial $U-c'(B_{0})$, if $T_{{\rm high}}$ is the number
of high-stage price updates and $D_{{\rm del}}$ the total number
of deleted elements, we have
\[
T_{{\rm high}}\le2\Gamma/K,\qquad D_{{\rm del}}=\sum_{j}b_{j}\le\sum_{j}(k_{j}-\delta_{j})\le\Gamma.
\]
Consider any point during the execution of the bit scaling phase.
We can bound the number of successful augmentations, including those
in the low stage, by 
\[
A_{{\rm total}}=|X|-|X_{0}|+D_{{\rm del}}\le\Gamma+D_{{\rm del}}\le2\Gamma.
\]
Let an interval be a portion of the high-deficit stage consisting
of a sequence of augmentations where the prices $p,q$ are constant.
Each interval uses $O(n(a+d)\log(r+1))$ queries, where $a$ is the
number of augmentations in the interval. The greedy basis changes
and all basis refreshes cost $O(n)$ queries each. There are at most
$T_{{\rm high}}+1$ such intervals. Putting everything together, we
obtain the following upper bound on the number of queries in the high-deficit
stage:
\begin{equation}
Q_{{\rm high}}=O\left(n\log(r+1)\left[\Gamma+\frac{r}{K}\left(1+\frac{\Gamma}{K}\right)\right]\right).\label{eq:near-nr-high}
\end{equation}

\subsection{Analysis of the number of queries in the low-deficit stage and the
overall algorithm}

In this section, we upper bound the number of queries in the low-deficit
stage. By combining with the bound for the high-deficit stage from
the previous section, we obtain an upper bound on the number of queries
for a scaling phase and the overall algorithm.

Let $\Gamma_{{\rm low}}=U_{{\rm entry}}-c'(B_{0})\le\Gamma$, where
$U_{{\rm entry}}$ is the dual value at the start of the low-deficit
stage. The size $|X|$ is now monotone. Index the executed epochs
by their distinct deficits $1\le k\le K$, and let $t_{k}$ be their
numbers of price updates. The summation over $k$ in this section
ranges over executed epochs only. Let $Q_{{\rm low}}$ denote the
number of queries in the low-deficit stage. A price update reduces
$U$ by exactly $k$, and thus 
\[
\sum_{k\le K}kt_{k}\le\Gamma_{{\rm low}}\le\Gamma.
\]
Define the harmonic number 
\[
H_{K}=\sum^{K}_{k=1}\frac{1}{k}\le1+\log K\le2\log(K+1).
\]
Proposition \ref{prop:multicut-epoch} gives, for an absolute constant
$C'$, 
\begin{align}
\mathbb{E}[Q_{{\rm low}}] & \le C'n\log(n+1)\,\mathbb{E}\left[\sum_{k\le K}\sqrt{r(t_{k}+1)}\right]\nonumber \\
 & =O\left(n\log(n+1)\left(\sqrt{r}\,K+\sqrt{r\Gamma H_{K}}\right)\right)\nonumber \\
 & =O\left(n\log(n+1)\left(\sqrt{r}\,K+\sqrt{r\Gamma\log(K+1)}\right)\right).\label{eq:near-nr-low}
\end{align}
The bound on the sum inside the expectation holds for every realized
history. Indeed, use $\sqrt{t_{k}+1}\le1+\sqrt{t_{k}}$ and weighted
Cauchy--Schwarz to obtain 
\[
\sum_{k\le K}\sqrt{r(t_{k}+1)}\le\sqrt{r}\,K+\sqrt{r\left(\sum_{k\le K}kt_{k}\right)\left(\sum_{k\le K}\frac{1}{k}\right)}\le\sqrt{r}\,K+\sqrt{r\Gamma H_{K}}.
\]
The factor $\sqrt{\log(K+1)}$ comes from $H_{K}$.

Combining with the upper bound on the number of queries $Q_{{\rm high}}$
in the high-deficit stage, we obtain the following query bound for
the scaling phase.
\begin{prop}
Expected cost of one scaling phase \label{prop:near-nr-phase} The
number of original independence queries in a bit scaling phase is
$O(nr\log(n+1)\sqrt{\log(r+1)})$ in expectation. The algorithm is
always correct.
\end{prop}

\begin{proof}
Take $K=\lceil\sqrt{r}\rceil$. Since $\Gamma\le r$, (\ref{eq:near-nr-high})
gives $Q_{{\rm high}}=O(nr\log(r+1))$. Also $\sqrt{r}\,K=O(r)$ and
$\sqrt{r\Gamma\log(K+1)}=O(r\sqrt{\log(r+1)})$, so (\ref{eq:near-nr-low})
gives 
\[
\mathbb{E}Q_{{\rm low}}=O(nr\log(n+1)\sqrt{\log(r+1)}).
\]
This absorbs the high-stage cost and the $O(n)$ initialization cost.
All price updates preserve the exact split $p+q=c'$ and maintain
a common independent set in the two current price-optimal matroids
$N_{1}(p)$ and $N_{2}(q)$. Every high-stage price update strictly
decreases the nonnegative gap $U(p,q)-c'(B_{0})$, and the number
of deletions and augmentations is bounded as above. In the  low stage,
each update either augments and increases $|X|$ or decreases the
gap $U(p,q)-c'(B_{0})$, ensuring termination. At $|X|=r$, the common
basis is maximum for both split prices and is therefore an exact optimum
for $c'$. Randomness changes only the query count, never the correctness
of an augmenting path or a price update.
\end{proof}

Finally, we have the following bound on the number of queries of the
overall bit scaling algorithm (Section \ref{sec:main-algorithm}).
Starting from the dummy common basis with zero split prices, the algorithm
performs $\lceil\log_{2}(W+1)\rceil$ bit scaling phases. Proposition~\ref{prop:near-nr-phase}
applies conditionally on the history at every phase. Summing conditional
expectations proves Theorem~\ref{thm:main} with total query bound
\[
O(nr\log(n+1)\sqrt{\log(r+1)}\log(W+1)).
\]
The logarithms have distinct sources: $\log(r+1)$ in the high-deficit
stage comes from neighbor binary search; the low-deficit stage search
costs are bounded using $\log(n+1)$, which also covers sampling repetitions;
$\sqrt{\log(r+1)}$ comes from the harmonic sum over deficits; and
$\log(W+1)$ counts bit scaling phases.

 \bibliographystyle{alpha}
\bibliography{references}
\end{document}